\documentclass[aps,prd,showpacs,twocolumn,superscriptaddress, floatfix,longbibliography]{revtex4-2}
\usepackage[utf8]{inputenc}
\usepackage{url}
\usepackage{multirow}
\usepackage{capt-of}
\usepackage{amsmath}
\usepackage{amssymb}
\usepackage{graphicx}
\usepackage{subfig}
\usepackage{parskip}
\usepackage{tikz}
\usepackage{verbatim}
\usepackage{rotating}
\usepackage{mathrsfs} 
\usepackage{float}
\usepackage{bm}
\usepackage{physics}
\usepackage{mathtools}
\usepackage{qcircuit}
\usepackage{array}
\usepackage{setspace}
\usepackage{ulem}
\usepackage{newfloat,algcompatible}
\usepackage[size=small]{caption}
\usepackage{etoolbox}
\AtEndEnvironment{algorithm}{\noindent\hrulefill\par\nobreak\vskip-8pt}

\DeclareFloatingEnvironment[
    fileext=loa,
    listname=List of Algorithms,
    name=\textbf{Strategy},
    placement=tbhp,
]{algorithm}
\DeclareCaptionFormat{algorithms}{\vskip-15pt\hrulefill\par#1#2#3\vskip-6pt\hrulefill}
\algblock[Input]{Input}{EndInput}
\algblockdefx[Input]{Input}{EndInput}%
    [1]{\textbf{Input} #1}%
    {}

\usepackage[unicode=true,bookmarks=true,bookmarksnumbered=false,bookmarksopen=false,
 breaklinks=false,pdfborder={0 0 1},backref=false,colorlinks=true]
 {hyperref}
 
\hypersetup{
 linkcolor=magenta, urlcolor=blue, citecolor=blue, pdfstartview={FitH}, unicode=true}

\newcommand{\sust}{Department of Physics, Southern University of Science and Technology, Shenzhen 518055, China}
\newcommand{\IQSE}{Shenzhen Institute for Quantum Science and Engineering, Southern University of Science and Technology, Shenzhen 518055, China}
\newcommand{\IQA}{International Quantum Academy, Shenzhen 518048, China}
\newcommand{\GKEY}{Guangdong Provincial Key Laboratory of Quantum Science and Engineering, Southern University of Science and Technology, Shenzhen 518055, China}
\newcommand{\SKEY}{Shenzhen Key Laboratory of Quantum Science and Engineering, Southern University of Science and Technology, Shenzhen 518055, China}

\begin{document}
\title{Simulating the Schwinger Model with a Regularized Variational Quantum Imaginary Time Evolution}

\author{Xiao-Wei Li}
\affiliation{\IQSE}

\author{Fei Li}
\affiliation{\sust}

\author{Jiapei Zhuang}
\email{zhuangjp@sustech.edu.cn}
\affiliation{\sust}
\affiliation{\IQSE}
\affiliation{\IQA}

\author{Man-Hong Yung}
\email{ yung@sustech.edu.cn}
\affiliation{\IQSE}
\affiliation{\IQA}
\affiliation{\GKEY}
\affiliation{\SKEY}

\date{\today}
\begin{abstract}
The Schwinger model serves as a benchmark for testing non-perturbative algorithms in quantum chromodynamics (QCD), emphasizing its similarities to QCD in strong coupling regimes, primarily due to the phenomena such as confinement and charge screening. However, classical algorithms encounter challenges when simulating the Schwinger model, such as the ``sign problem"  and the difficulty in handling large-scale systems. These limitations motivate the exploration of alternative simulation approaches, including quantum computing techniques, to overcome the obstacles. While existing variational quantum algorithms (VQAs) methods for simulating the Schwinger model primarily rely on mathematical gradient-based optimization, which sometimes fail to provide intuitive and physically-guided optimization pathways. In contrast, the Variational Quantum Imaginary Time Evolution (VQITE) method offers a physically-inspired optimization approach. Therefore, we introduce that VQITE holds promise as a potent tool for simulating the Schwinger model. However, the standard VQITE method is not sufficiently stable, as it encounters difficulties with the non-invertible matrix problem. To address this issue, we have proposed a regularized version of the VQITE, which we have named the Regularized-VQITE (rVQITE) method, as it incorporates a truncation-based approach. Through numerical simulations, we demonstrate that our proposed rVQITE approach achieves better performance and exhibits faster convergence compared to other related techniques. We employ the rVQITE method to simulate the phase diagrams of various physical observables in the Schwinger model, and the resulting phase boundaries are in agreement with those obtained from an exact computational approach.
\end{abstract}

\maketitle

\section{Introduction}
As an important application scenario of quantum algorithms in the NISQ era, quantum simulation of quantum field theory has attracted increasing interest~\cite{martinez2016real, kokail2019self, yang2020simulating, yang2020observation, atas20212su, chakraborty2022classically,nagano2023quench, ikeda2023detecting}. 
The Schwinger model is a Quantum Electrodynamics (QED) model defined in $(1+1)$-dimensional Minkowski spacetime~\cite{schwinger1962gauge}. 
Despite its relative simplicity and the existence of analytical solutions in the massless limit~\cite{schwinger1962gauge,lowenstein1971quantum}, it remains a worthwhile field theory model due to its manifestation of Quantum Chromodynamics (QCD)-like properties, such as confinement and charge screening, under strong coupling~\cite{coleman1975charge,gross1996screening}. 
These characteristics make it an ideal platform for exploring novel algorithms for QCD in four-dimensional spacetime~\cite{ikeda2023detecting}. However, classical algorithms encounter certain challenges when simulating the Schwinger model. For instance, the notorious "sign problem" can hinder the efficiency of numerical methods such as Monte Carlo method~\cite{troyer2005computational,fukushima2010phase,nagata2022finite}. These limitations motivate the exploration of alternative simulation approaches, including quantum computing techniques, to overcome the obstacles presented by the Schwinger model.

The quantum simulations of the Schwinger model are mainly based on the lattice representation of the Schwinger model Hamiltonian~\cite{kogut1975hamiltonian}. Previous research has experimentally simulated the Schwinger model in a 71-site ultra-cold atomic optical lattice system ~\cite{yang2020observation}.   The Schwinger model has also been studied using variational quantum algorithms (VQA), in which the lattice Schwinger Hamiltonian is mapped onto a spin model via the Jordan-Wigner transformation.
A recent work employed the $\beta$-VQE to investigate the phase diagram of the Schwinger model at finite temperature and density ~\cite{tomiya2022schwinger}. Furthermore, another study utilized Variational Quantum Simulation (VQS) to explore the Schwinger model in the presence of an external electric field~\cite{nagano2023quench}.

While previous works~\cite{tomiya2022schwinger,nagano2023quench} have attempted to simulate the Schwinger model using VQA algorithms, these VQA approaches primarily rely on mathematical gradient descent optimization, which can at times fail to provide intuitive and instructive optimization pathways. 
In this work, we discuss an imaginary-time evolution method~\cite{mcardle2019variational, motta2020determining} that selectively extracts the ground state component by following a physically principled optimization trajectory. However, the imaginary-time evolution operator is non-unitary, presenting a challenge for implementation on quantum computers, which require unitary operations. To overcome this limitation, several methods have been proposed, including quantum imaginary-time evolution (QITE) ~\cite{motta2020determining, yeter2022quantum},  variational quantum imaginary-time evolution (VQITE) ~\cite{Yuan2019theoryofvariational,mcardle2019variational}, and probabilistic imaginary-time evolution (PITE) ~\cite{liu2021probabilistic,kosugi2021probabilistic,xie2024probabilistic}.

In our work, we primarily utilize the powerful tools of variational quantum imaginary-time evolution (VQITE) with McLachlan's variational principle. However, upon analyzing the standard VQITE algorithm workflow~\cite{Yuan2019theoryofvariational,mcardle2019variational}, we identified that the parameter update step encounters a numerical stability issue. To address this, we have proposed a regularized VQITE (rVQITE) protocol. Through this revised approach, we present a comprehensive study of the Schwinger model in the presence of an external field and at finite chemical potential and zero temperature. At zero temperature, the system's ground state is sought by minimizing the expectation value of the Hamiltonian. 
The ansatz employed for this purpose is the Hamiltonian Variational Ansatz (HVA) ~\cite{wiersema2020exploring}, a hybrid quantum-classical algorithm that iteratively refines the wavefunction by tuning variational parameters. 

To validate the precision and effectiveness of our rVQITE approach, we compare our results with those obtained from the exact solution of the Schwinger model at zero chemical potential and zero external fields. This comparison demonstrates the high level of accuracy achieved by our method. 
Additionally, By carefully adjusting the initial conditions, we studied the behavior of the system across different combinations of mass, chemical potential, and external fields. We constructed phase diagrams that included the $U(1)$ charge, chiral condensate, and total electric field, where we observed distinct phase transition phenomena.
We phenomenologically demonstrated the quasi-symmetric/quasi-antisymmetric behaviors of observables under mirror transformations for different parameter combinations.
To elucidate the underlying mechanisms of these phase transitions, we examined the hierarchical structure of the energy levels associated with the lowest charged states as a function of charge quantity. This analysis revealed the critical role of the chemical potential in mediating the transitions through these hierarchical energy structures.

The paper is organized as follows. In Sec.~\ref{sec:lattice}, we introduce the Schwinger model, its spin representation, and some observables of interest. In Sec.~\ref{sec:vqite_ground}, we first present the ansatz used for state preparation. We primarily describe the variational quantum imaginary-time evolution (VQITE) method and propose our improved version, the Regularized-VQITE (rVQITE). In Sec.~\ref{sec:simulations}, we benchmark the performance of our proposed rVQITE approach and use it to simulate the phase diagram of the Schwinger model. Finally, we give conclusions in Sec.~\ref{sec:conclusion}.

\section{Background}
\subsection{The Schwinger model and its spin representation}\label{sec:lattice}

The Schwinger model constitutes a quantum electrodynamics (QED) framework delineated within the context of $(1+1)$-dimensional Minkowski spacetime. This spacetime manifold is characterized by coordinates $\left(x^0, x^1\right)$, endowed with a Minkowski metric $\eta_{\mu\nu} = \operatorname{diag}(+1, -1)$, which facilitates the description of relativistic phenomena in two dimensions. Notably, the model exhibits striking parallels with quantum chromodynamics (QCD), most notably in its manifestation of confinement (the massive case) and the chiral symmetry breaking. These features render the Schwinger model a quintessential testbed for algorithmic developments aimed at addressing challenges in lattice gauge theory, thereby providing valuable insights into non-perturbative aspects of gauge theories.

In our analysis, we embark upon the formulation of the target Hamiltonian for the Schwinger model augmented with a topological term, cast in the language of spin representation. The intricate derivation of this Hamiltonian from its Lagrangian counterpart is meticulously detailed in Appendix~\ref{appendix:a}, ensuring a comprehensive understanding for the interested reader. The resulting target Hamiltonian, expressed in the spin basis, is presented below:

\begin{equation}
\begin{aligned}
H(m,\theta) = & J \sum_{j=0}^{N-2}\left[\sum_{i=0}^j \frac{\sigma_i^z + (-1)^i}{2} + \frac{\theta}{2\pi}\right]^2 \\
&+ \frac{w}{2} \sum_{j=0}^{N-2}\left[\sigma_j^x \sigma_{j+1}^x + \sigma_j^y \sigma_{j+1}^y\right] \\
&+ \frac{m}{2} \sum_{j=0}^{N-1}(-1)^j \sigma_j^z,
\end{aligned}
\end{equation}
wherein the parameters $w = 1/(2a)$ and $J = g^2a/2$ encapsulate the fundamental constants of the system, with $a$ denoting the lattice spacing and $g$ representing the coupling strength. As mentioned in Ref.~\cite{nagano2023quench}, the term $\theta/(2\pi)$ serves as an external field.

To delve into the Schwinger model under conditions of finite density, we incorporate the chemical potential $\mu$ into the Hamiltonian through the grand canonical ensemble, effectively modifying the time derivative component according to:
\begin{equation}
    i\partial_0 \rightarrow i\partial_0 - \mu,
\end{equation}
yielding the Hamiltonian in the presence of finite density:
\begin{equation}
    H(\mu, m, \theta) = H(m, \theta) - \mu Q,
\end{equation}
where $Q = \frac{1}{2}\sum_{j=0}^{N-1}\sigma_j^z$ serves as the $U(1)$ charge operator (see Appendix~\ref{appendix:a}).

In our study of the Schwinger model, the quantities of primary interest emerge as the expectation values of pertinent operators, denoted as $\langle O \rangle = \bra{\psi} O \ket{\psi}$, with $\ket{\psi}$ representing the ground state of $H(\mu, m, \theta)$. Specifically, the total electric field is calculated as follows:
\begin{equation}
\begin{aligned}
\mathcal{E} &= \frac{g}{N} \sum_{i=0}^{N-1} \left \langle L_i + \frac{\theta}{2\pi} \right\rangle \\
&= \frac{g}{2N} \sum_{i=0}^{N-1} \sum_{k=0}^i \left( \langle \sigma_k^z \rangle + (-1)^k \right) + \frac{g\theta}{2\pi}.
\end{aligned}
\end{equation}
The chiral condensate is given by:
\begin{equation}
    \chi = \frac{\langle \bar{\psi} \psi \rangle}{Na} = \frac{ag}{2N} \sum_{i=0}^{N-1} (-1)^i \langle \sigma_i^z \rangle.
\end{equation}
And the average $U(1)$ charge is determined by:
\begin{equation}
    \langle Q \rangle = \frac{1}{2} \sum_{i=0}^{N-1} \langle \sigma_i^z \rangle.
\end{equation}

A salient characteristic of the model is the commutation of the $U(1)$ charge operator $Q$ with the Hamiltonian $H(0, m, \theta)$, i.e., $[H(0,m,\theta), Q]=0$. This is a property that underscores the conservation of charge throughout the system's evolution. This symmetry permits the eigenstates of $H(0, m, \theta)$ to be distinguished by their associated charge and energy levels, denoted as $\ket{q, n}$, where $q$ signifies the $U(1)$ charge and $n$ represents the energy level for a fixed charge number, with $n=0$ signifying the ground state or lowest energy level.
The eigenfunction equations of Hamiltonian and $U(1)$ charge are $H(0, m, \theta) \ket{q, n} = E_n^{(q)}(0,m,\theta) \ket{q, n}$,
$Q \ket{q, n} = q \ket{q, n}$.

\subsection{Variational quantum imaginary-time evolution}\label{sec:vqite}

Quantum Imaginary-Time Evolution (QITE) is a promising method to prepare the ground state in quantum computers~\cite {motta2020determining}. It allows correlations to build faster than would
be allowed by the Lieb-Robinson bond that governs real-time
evolution~\cite{bench2019making}, and always converges to ground state~\cite{motta2020determining}.
The QITE is based on the imaginary-time
Schr\"{o}dinger equation
\begin{equation}\label{eq:dynamic}
    \frac{d}{d\tau }\ket{\psi(\tau)}=-(H-E_\tau)\ket{\psi(\tau)},
\end{equation}
where $E_\tau=\langle\psi(\tau)|H| \psi(\tau)\rangle$, the state at imaginary time $\tau$ with normalised condition is $|\psi(\tau)\rangle:=$ $e^{-H \tau}|\psi(0)\rangle / \sqrt{\left\langle\psi(0)\left|e^{-2 H \tau}\right| \psi(0)\right\rangle}$.
The ground state of the system Hamiltonian $H$ can be represented as the long-time limit of this imaginary time state, $\lim _{\tau \rightarrow \infty} \frac{|\psi(\tau)\rangle}{\|\ket{\psi(\tau)}\|}$. 
However, QITE suffers a circuit depth growing exponentially with the correlation domain sizes (being
roughly the system’s correlation length) and linearly with
the number of imaginary time steps~\cite{gomes2021adaptive}.
An alternative approach is to combine the QITE with variational quantum eigensolver (VQE) that is based on an energy cost function of a fixed depth variational ansatz that to be minimized-Variational Quantum Imaginary-Time Evolution(VQITE)~\cite{Yuan2019theoryofvariational,mcardle2019variational}. This approach can be understood as a special VQE case with quantum natural gradient optimization~\cite{stokes2020quantumnatural}.

Let us briefly recap the fundamental framework of VQITE.
Consider a parameterized state $\ket{\psi(\bm{\theta}(\tau))}$ with real and $\tau$-dependent parameters $\bm{\theta}$, the dynamic equation Eq.~\eqref{eq:dynamic} becomes
\begin{equation}\label{eq:para_dynamic}
    \left[\sum_i \dot{\theta}_i\frac{\partial }{\partial \theta_i}+(H-E_\tau)\right]\ket{\psi(\bm{\theta})}=\bm{0}. 
\end{equation}

The imaginary-time evolution can be simulated by the parameters update rule with a small time difference $\delta \tau$
\begin{equation}
   \bm{\theta}(\tau +\delta \tau)\approx\bm{\theta}(\tau)+\delta\tau \dot{\bm{\theta}}.
\end{equation}
Therefore, the key step of imaginary-time evolution is to determine the imaginary time derivative of parameters $\dot{\bm{\theta}}$. 

Following McLachlan’s variational principle
\begin{equation}\label{eq:mclachlan}
    \delta \left |\left| (\frac{d}{d\tau} + H - E_\tau)\ket{\psi(\tau)}\right |\right|=0, 
\end{equation}
the evolution with real parameters becomes~\cite{Yuan2019theoryofvariational} 
\begin{equation}\label{eq:evolution}
    A^R \dot{\bm{\theta}}=-C^R,
\end{equation}
where 
\begin{align}
\left(A^R\right)_{i j} &=\operatorname{Re}\left[\frac{\partial\langle\psi(\boldsymbol{\theta}(\tau))|}{\partial \theta_i} \frac{\partial|\psi(\boldsymbol{\theta}(\tau))\rangle}{\partial \theta_j}\right],\\
\left(C^R\right)_i&=\operatorname{Re}\left[\frac{\partial\langle\psi(\boldsymbol{\theta}(\tau))|}{\partial \theta_i} H|\psi(\boldsymbol{\theta}(\tau))\rangle\right].
\end{align}
Then, the imaginary time derivative of parameters is given by 
\begin{equation}
    \dot{\bm{\theta}}=-(A^R)^{-1} C^R.
\end{equation}

\section{Method}\label{sec:vqite_ground}

\subsection{The ansatz for state preparation}
In the previous section, we presented the spin representation of the Schwinger model. We leverage the Hamiltonian Variational Ansatz (HVA) within this context to articulate the quantum state, as delineated in Ref.~\cite{wiersema2020exploring}:
\begin{widetext}
\begin{equation}
\ket{\psi(\bm{\alpha},\bm{\beta},\bm{\gamma})} = \prod_{l=1}^p\left[U_{xy,even}(\bm{\beta}_l)U_{zz,even}(\bm{\gamma}_l) U_{xy,odd}(\bm{\beta}_l)U_{zz,odd}(\bm{\gamma}_l)U_z(\bm{\alpha}_l)\right]\ket{\psi_{\text{in}}},
\end{equation}
\end{widetext}
where the individual components of the ansatz are defined as:
\begin{equation}
\begin{aligned}
U_{xy,even}(\bm{\beta}_l) &:= \prod_{n:even}\exp[i\beta_{l,n}(\sigma_n^x\sigma_{n+1}^x+\sigma_n^y\sigma_{n+1}^y)], \\
U_{zz,even}(\bm{\gamma}_l) &:= \prod_{n:even}\exp[i\gamma_{l,n}\sigma_n^z\sigma_{n+1}^z], \\
U_{xy,odd}(\bm{\beta}_l) &:= \prod_{n:odd}\exp[i\beta_{l,n}(\sigma_n^x\sigma_{n+1}^x+\sigma_n^y\sigma_{n+1}^y)], \\
U_{zz,odd}(\bm{\gamma}_l) &:= \prod_{n:odd}\exp[i\gamma_{l,n}\sigma_n^z\sigma_{n+1}^z], \\
U_{z}(\bm{\alpha}_l) &:= \prod_{n=0}^{N-1}\exp[i\alpha_{l,n}\sigma_n^z].
\end{aligned}
\end{equation}

It is imperative to note that the $U(1)$ charge  remains invariant under variations of the parameters $(\bm{\alpha},\bm{\beta},\bm{\gamma})$. Instead, its value is entirely determined by the initial setup encapsulated within $\ket{\psi_{\text{in}}}$. 

For situations where $q$ is fixed, we set the first $2|q|$ qubit to manifest a charge of $q$, with the rest of the qubits being configured in the bare vacuum state. Precisely, the bare vacuum state is expressed as $\ket{0101\cdots01}$, signifying a state without any fermion or antifermion excitations. When $q > 0$, the first $2|q|$ qubits are prepared in the $\ket{0}$ state; in contrast, when $q < 0$, they are initialized to $\ket{1}$. In instances where $q$ is not predetermined, we incorporate $N$ parameterized rotation gates $R_{x}(\tau_i)=e^{-i\tau_i\sigma_i^x}$ into the initial state. These parameters are involved in the variational imaginary time evolution, enabling us to tune to the correct $q$ value.

\begin{figure*}[htp]
\centering
\includegraphics[width=1.0\textwidth]{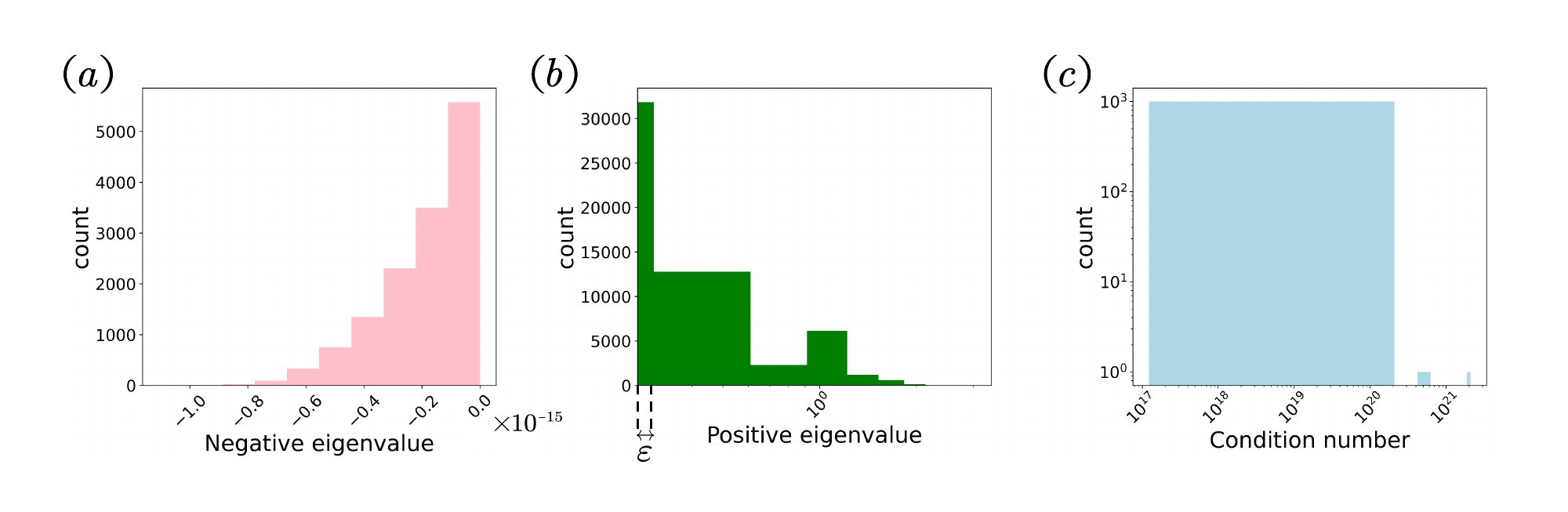}
\caption{(a)The distribution of negative eigenvalues of matrix $A^R$. (b)The distribution of positive eigenvalues of matrix $A^R$ (c)The distribution of condition numbers for matrix $A^R$. The distributions presented were obtained through random sampling of 10 instances with system size $N=10$, simulated using the standard VQITE approach applied to the Schwinger model.}
\label{Fig:eigenvalue spectrum}
\end{figure*}

\subsection{Regularized variational quantum imaginary-time evolution}\label{sec:mvqite}
Herein, when executing the VQITE algorithm, we employ the squared McLachlan distance~\cite{Yuan2019theoryofvariational, gomes2021adaptive} defined as
\begin{equation}
\begin{aligned}
\label{eq_error}
    \Delta^2 &= \left|\left|\left[\sum_i \dot{\theta}_i\frac{\partial }{\partial \theta_i}+(H-E_\tau)\right]\ket{\psi(\bm{\theta})}\right|\right|^2\\
    &=\dot{\bm{\theta}}^\top A^R \dot{\bm{\theta}}+2\dot{\bm{\theta}}^\top C^R +\text{Var}_{\bm{\theta}} (H)
\end{aligned}
\end{equation}
to justify whether the imaginary time Schr\"{o}dinger equation is satisfied.
Here $||\ket{\psi}||=\sqrt{\langle\psi\ket{\psi}}$ is the Frobenius norm of quantum state, and $\text{Var}_{\bm{\theta}} (H) =\bra{\psi(\bm{\theta})}H^2\ket{\psi(\bm{\theta})}-E_\tau^2$.
The squared McLachlan distance reflects how close the evolution path is to a strict imaginary time evolution path, that is, the smaller the error, the more the evolution path satisfies the imaginary time Schr\"{o}dinger equation.

Notably, the parameter update rule embedded in Equation~\eqref{eq:evolution} fundamentally represents a quadratic minimization problem. This rule is, in essence, synonymous with the critical point condition of the minimization problem associated with the squared McLachlan distance, where the variable vector of interest is $\dot{\bm{\theta}}$. Explicitly, this equivalence is captured by the following expression:
\begin{equation}
    \frac{\partial}{\partial \dot{\bm{\theta}}} \left[\dot{\bm{\theta}}^\top A^R \dot{\bm{\theta}} + 2\dot{\bm{\theta}}^\top C^R + \text{Var}_{\bm{\theta}}(H)\right] = 0.
\end{equation}
Given that the norm of the state vector is inherently constrained to non-negative real numbers, the nature of the extremum identified through this process is assuredly a minimum.

The positive definiteness of $A^R$ is the key point to ensure numerical stability. The function $\Delta^2(\dot{\bm{\theta}})$ opens upwards in every direction of the variable space, which is a fundamental requirement for establishing the existence of a global minimum.
Without the guarantee of positive definiteness of $A^R$, one cannot assert the presence of a global minimum with certainty. 
However, no evidence guarantees the positive definiteness of $A^R$. 
Furthermore, even if $A^R$ is positively defined, the situation that $A^R$ exists very small eigenvalues ($\lambda_i\ll 1$) can also hinder the search for the minimum value point. 
Thus, to ensure the stability of the imaginary time evolution algorithm, it is essential to regularize matrix $A^R$ thereby ensuring its favorable properties for reliable computations.

To investigate the numerical stability of matrix $A^R$, we studied its eigenvalue distribution.
Fig.~\ref{Fig:eigenvalue spectrum}~(a) and (b) display the eigenvalue distribution of the matrix $A^R$, which was employed in the simulation of the Schwinger model. The results show that a significant number of the eigenvalues are concentrated near 0. 
Especially,  a significant number of negative eigenvalues that cannot be ignored, as shown in Fig.~\ref{Fig:eigenvalue spectrum}~(a), 
which violates the positive-definiteness requirement. 
The near-zero eigenvalues lead to a small determinant of the matrix $A^R$, $\operatorname{det}(A^R)=\prod_{i=1}^n \lambda_i$, to the extent that it is numerically considered as 0. This is one of the factors contributing to the non-invertibility of the matrix $A^R$. We also computed the condition number $\kappa(A^R)=\left|\frac{\lambda_{\max }}{\lambda_{\min }}\right|$  of the matrix  $A^R$ in Fig.~\ref{Fig:eigenvalue spectrum} (c), where $\lambda_{\max }$ and $\lambda_{\min }$ denote the largest and smallest singular values of $A^R$, respectively. The condition number is a metric that can measure the sensitivity of the output value to perturbations in the input data and round off errors during the solution process~\cite{anguas2019comparison,el2002inversion}. A high condition number indicates that the matrix is ill-conditioned, making the numerical computations more prone to instability and inaccuracy. These observed results highlight the inherent instability associated with the inversion of the matrix \(A^R\), which poses a limiting factor in the practical application of the VQITE method.

We introduce a Regularized Variational Quantum Imaginary-Time Evolution (rVQITE) protocol to navigate this challenge. Our approach commences with diagonalizing $A^R$ using an orthogonal matrix $\Lambda$, yielding $\Lambda^\top A^R \Lambda = \text{diag}(\lambda_1, \ldots, \lambda_m)$, where $\lambda_i$ denotes the $i$-th eigenvalue of $A^R$, and $m$ represents the dimension of the parameter vector $\bm{\theta}$.
Under this orthogonal transformation, Eq.~\eqref{eq:evolution} becomes
\begin{equation}
    (\Lambda^\top A^R\Lambda) (\Lambda^\top \dot{\bm{\theta}})=-\Lambda^\top C^R.
\end{equation}
Consequently, the transformed parameter update 
\begin{equation}
\begin{aligned}
\bm{g} := \Lambda^\top\dot{\bm{\theta}}=-(\Lambda^\top A^R\Lambda)^{-1} (\Lambda^\top C^R).
\end{aligned}
\end{equation}

Next, we adopt a threshold \(\varepsilon\) to segregate eigenvalues, distinguishing `well-conditioned' $w$ from `ill-conditioned' $i$ dimensions based on whether their eigenvalues exceed $\varepsilon$. This permits a decomposition of the parameter update space into components \(\bm{g}_w\) and \(\bm{g}_i\), mirroring a similar partitioning of matrices derived from \(A^R\) and \(C^R\). Specifically, we have \(\Lambda^\top A^R \Lambda = A_w \oplus A_i\), \(\bm{g} = \bm{g}_w \oplus \bm{g}_i\), and \(\Lambda^\top C^R = \bm{c}_w \oplus \bm{c}_i\), employing \(\oplus\) to denote the direct sum. As examples, for matrices $B$, $D$ and vectors $\bm{u}$, $\bm{v}$, the direct sum are given by
\begin{equation}
    B\oplus D=
    \left[
        \begin{matrix}
            B & 0\\
            0 & D
        \end{matrix}
    \right],\quad
    \bm{u}\oplus \bm{v}=\left[
    \begin{matrix}
        \bm{u}\\
        \bm{v}
    \end{matrix}
    \right].
\end{equation}
To rectify `ill-conditioned' influences, we nullify the `ill-conditioned' component \(\bm{g}_i\), setting it to the zero vector \(\bm{0}_i\). Consequently, the refined parameter update formula becomes
\begin{equation}
\begin{aligned}
\dot{\bm{\theta}} = \Lambda(\bm{g}_w \oplus \bm{0}_i)
\end{aligned}
\end{equation}
with \(\bm{g}_w = - A_w^{-1} \bm{c}_w\), thereby ensuring a physically consistent and computationally viable parameter update scheme.

\section{Numerical results }\label{sec:simulations}
In this section, we present a quantitative assessment of the performance of the rVQITE algorithm through numerical simulations. We conduct a comparative analysis by benchmarking the rVQITE approach against other relevant methods. Furthermore, we demonstrate the application of the rVQITE algorithm in the context of phase diagram simulations for the Schwinger model. We utilize the Python library \textit{MindQuantum} \cite{mq_2021} to simulate the quantum circuits in classical computers.

\subsection{Algorithm performance and benchmarking }
We use the solution of the Schwinger model at zero chemical potential and zero external field as a benchmark to test our rVQITE method.
We utilize the dimensionless quantity `Ratio' as a quantitative metric to assess the accuracy of the algorithm~\cite{pagano2020quantum}.  
\begin{equation} 
\text { Ratio }:=\frac{\left(E_{\max }-E_{\mathrm{QA}}\right)}{\left(E_{\max }-E_{\min }\right)}
\end{equation}
where $E_{\max / \min }$ is the highest/lowest eigenvalue of the Hamiltonian obtained through exact methods. $E_{\mathrm{QA}}$ is ground state energy obtained by quantum algorithms. As the ratio value approaches 1, it indicates that the ground state obtained by the quantum algorithms is increasingly closer to the true ground state, thereby suggesting improved performance of the corresponding quantum algorithms.

\begin{figure}[htp]
\centering
\includegraphics[width=0.50\textwidth]{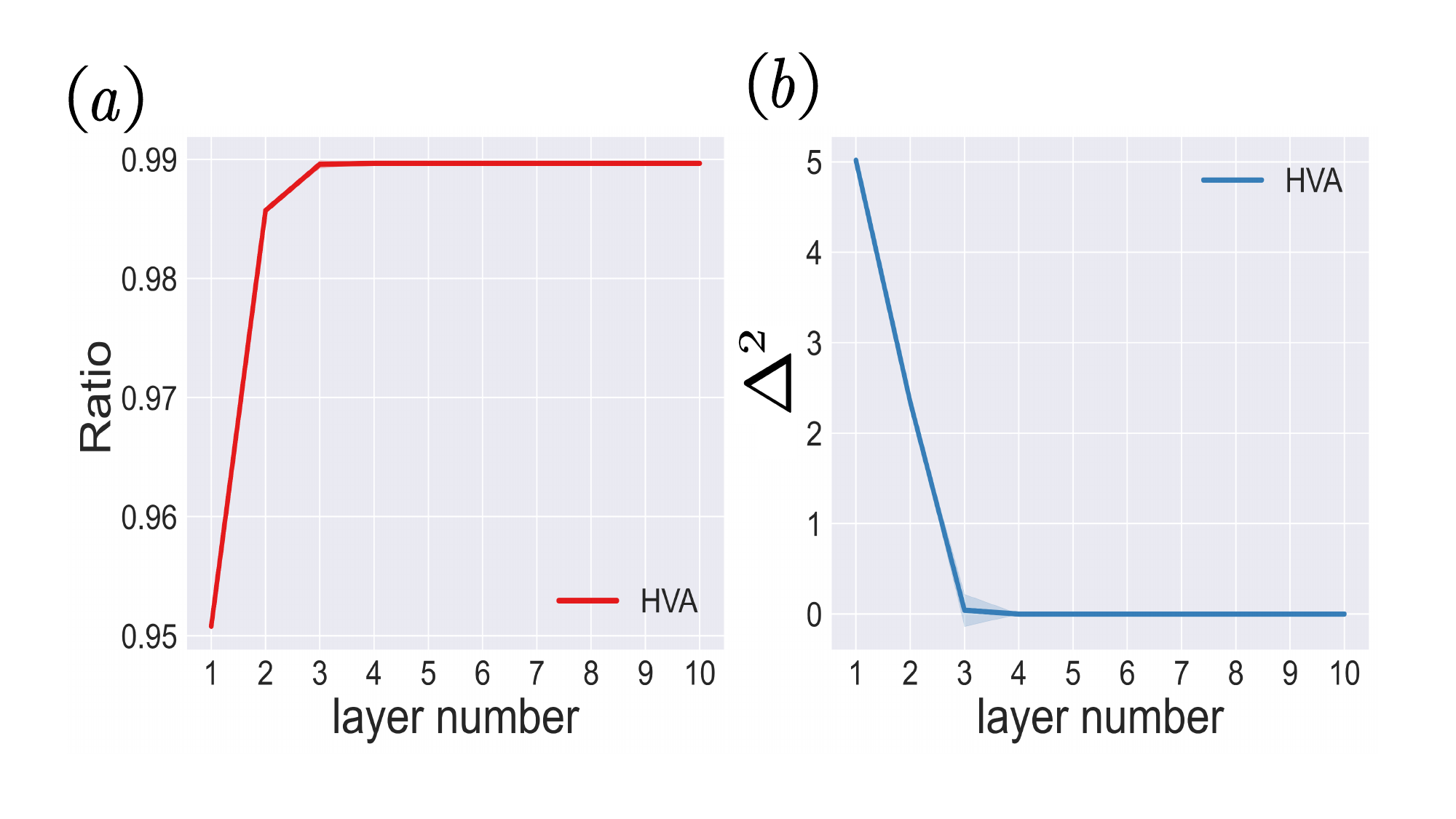}
\caption{(a)The ratio value of regularized variational quantum imaginary-time evolution for different HVA depth $p=1,2,3,4,5,6,7,8,9,10$. 
(b) The corresponding squred McLachlan distance $\Delta^2$ in (a).}
\label{Fig:distance}
\end{figure}

To evaluate the performance of the rVQITE algorithm, we conducted a series of tests examining the ratio values of rVQITE under varying depths of the Hamiltonian Variational Ansatz (HVA). For each HVA depth, we executed 20 samples with random initializations. In Fig. \ref{Fig:distance} (a), we present these evolutionary results' mean and standard deviation. Notably, the mean and standard deviation are difficult to distinguish due to the consistent performance after 500 iterations for each sample. Additionally, in Fig. \ref{Fig:distance} (b), we compute the square McLachlan distance $\Delta^2$ corresponding to the results shown in Fig.~ \ref{Fig:distance} (a), with the McLachlan distance calculation formula provided by Eq.~\eqref{eq_error}. The square McLachlan distance $\Delta^2$ quantifies the degree to which the proposed rVQITE method faithfully follows the desired trajectory along the imaginary time evolution. The numerical results of these two outcomes show that even with a single-layer HVA, the rVQITE ratio can reach 0.95, and as the iterative path becomes more faithful to the imaginary time evolution, the ratio can reach higher values. Furthermore, when the HVA depth exceeds 4 layers, the ratio converges to 0.99, with the square McLachlan distance approaching zero.

\begin{figure}[htp]
\centering
\includegraphics[width=0.45\textwidth]{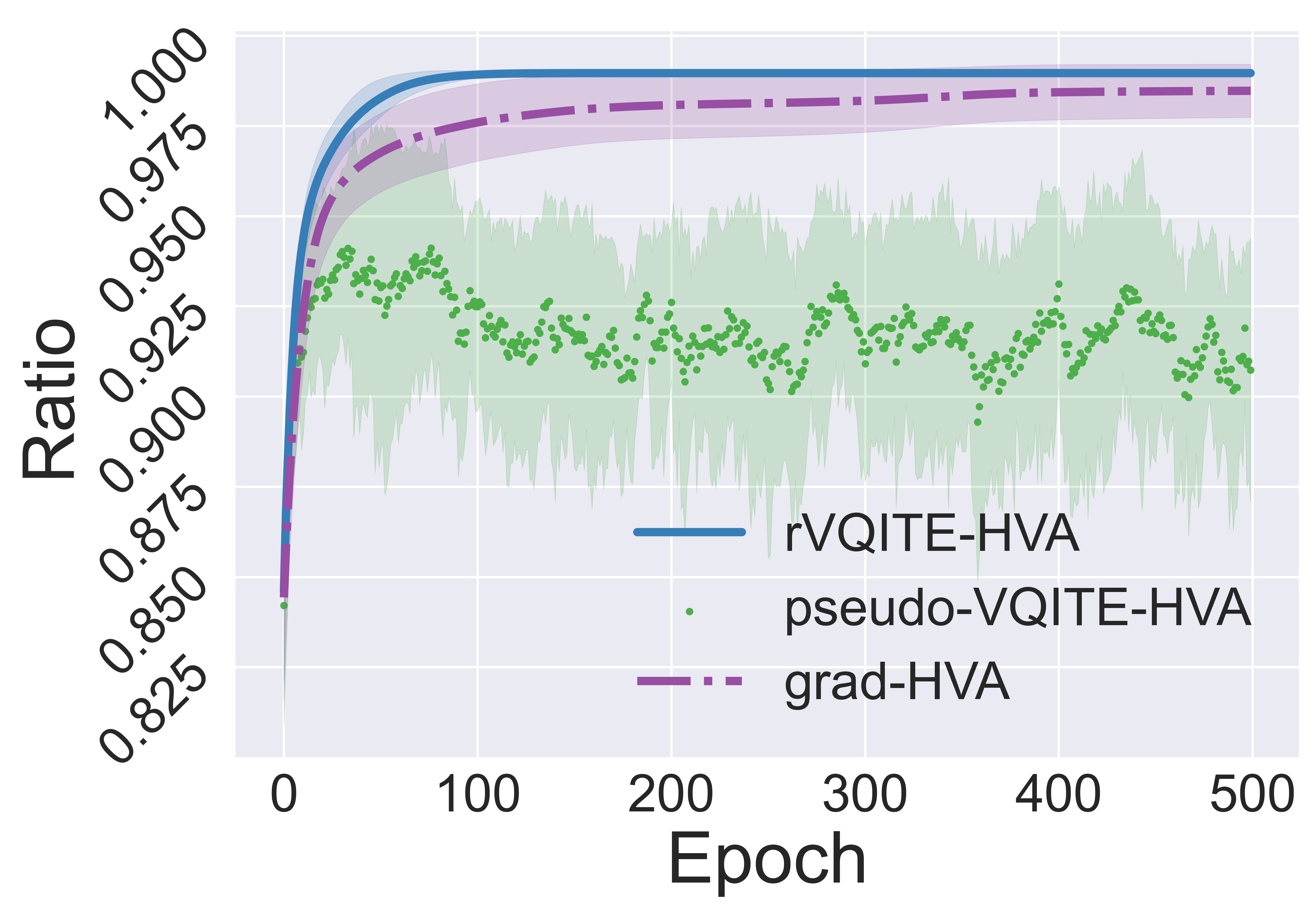}
\caption{A comparison of regularized variational quantum imaginary-time evolution(rVQIT), the pseudo-inverse-based Variational Quantum Imaginary Time Evolution (pseudo-VQITE), and gradient-based optimization Method. Each method's standard deviation and mean were computed for 20 randomly initialized samples.}
\label{Fig:mVQITE}
\end{figure}

In Fig. \ref{Fig:mVQITE}, we present a comparative analysis of the evolution of the Ratio values as a function of iteration count for three distinct methods: rVQITE, pseudo-inverse, and gradient descent. The figure displays the mean and standard deviation of the Ratio values, where each method was executed across 20 randomly initialized samples. 
This result demonstrates that our rVQITE method exhibits superior convergence speed compared to gradient-based optimization techniques. Moreover, when compared with the pseudo-inverse-based Variational Quantum Imaginary Time Evolution (pseudo-VQITE), our approach showcases enhanced stability, consistently converging to a ratio close to unity. This feature underscores the robustness and accuracy of our method, positioning it as a preferred choice for simulations requiring both precision and computational efficiency.

\subsection{The phase diagram of the Schwinger model}
\begin{figure*}[t]
    \centering
    \includegraphics[width=\textwidth]{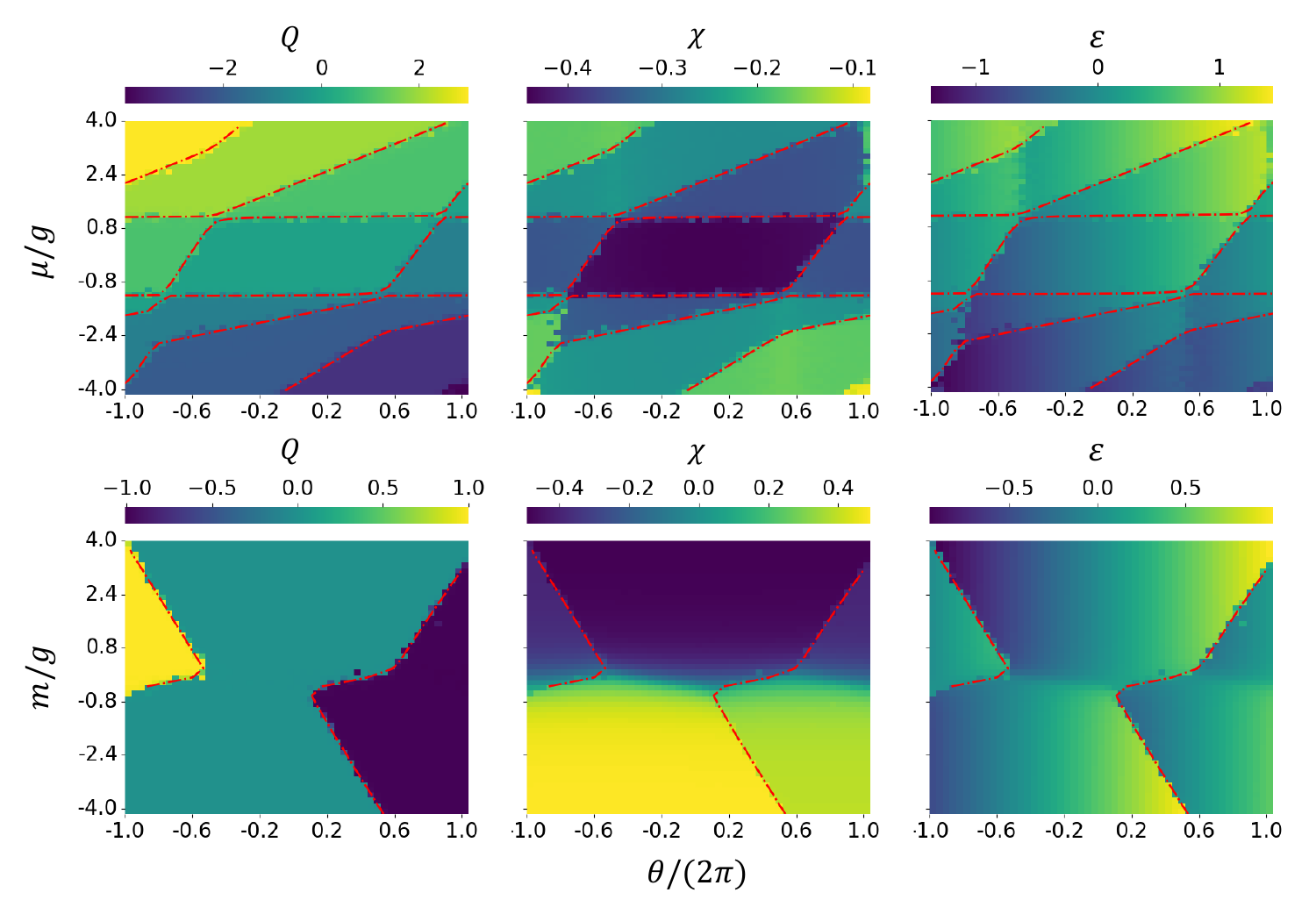}
    \caption{
    From left to right, heat maps depict the $U(1)$ charge, chiral condensate, and electric field, respectively. The top row presents these quantities for varying values of $\theta$ and chemical potential $\mu$, whereas the bottom row shows them for different values of $\theta$ and mass $m$. Throughout both rows, we set $N=10$ and $a=1/g$. In the top row, the mass is held constant at $m=g$, while in the bottom row, the chemical potential is fixed at $\mu=0$.
    The red dash-dot curves depict the numerical solutions of Eq.~\eqref{eq:phase_equation} under different parameter settings, where the energy values in the equation are obtained through exact diagonalization. The dash-dot curves are significantly consistent with the phase boundaries reflected in the heat maps, which not only attests to the accuracy of the rVQITE algorithm but also validates the physical essence encapsulated within the phase boundaries.}
    \label{fig:phase}
\end{figure*}

\begin{figure}[t]
    \centering
    \includegraphics[width=\linewidth]{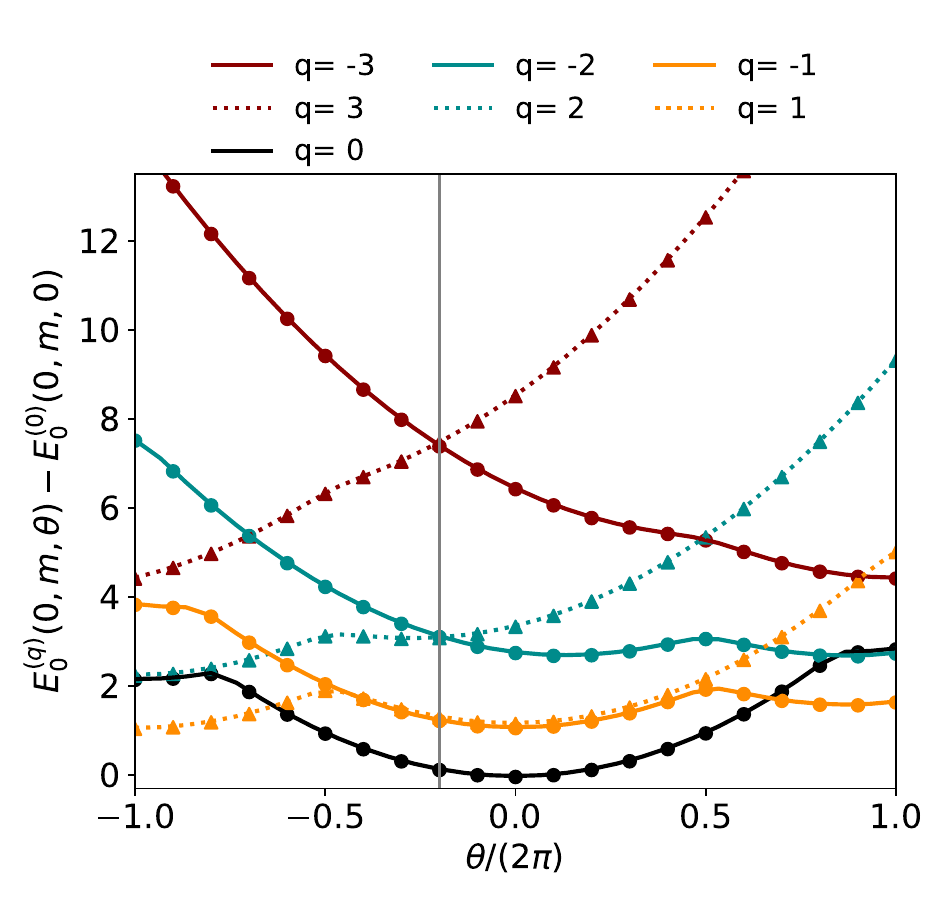}
    \caption{Under zero chemical potential, we investigated the relationship between the lowest energy level for a fixed charge $q$ and the angle $\theta$. Specifically, we set the initial state's charge to $q = -3, -2, -1, 0, 1, 2, 3$ and employed our rVQITE method to compute the corresponding lowest energies. The $x$-coordinate where the gray solid line intersects marks the point at which the lowest energy level for $q$ coincides with that for $-q$.  Here we set $N=10$, $m=g$, $\mu=0$ and $a=1/g$.
    The solid and dashed lines represent the results of exact diagonalization, while circles and triangles mark the outcomes of our rVQITE calculations. There is excellent agreement between the two.}
    \label{fig:energy}
\end{figure}

In this part, we employ the rVQITE method we proposed to simulate the phase diagrams of the Schwinger model. We compare the simulated phase diagram results with the phase boundaries obtained using the exact diagonalization-based roots-finding method, and the results demonstrate excellent
agreement between the two approaches.

Firstly, we investigate the behavior of $U(1)$ charge, chiral condensate, and electric field across varying chemical potential $\mu$ and $\theta$ angle, as depicted in Fig.~\ref{fig:phase}~(a), (b), and (c). For this purpose, we employ a system size of $N = 10$, with parameters set as $a = 1/g$ and $m = g$. Our approach utilizes a 5-layer HVA ansatz with an initial state of  $\prod_{i=1}^N R_{x}(\tau_i) \ket{00\cdots 0}$, in conjunction with our rVQITE method to prepare ground states and measure the expectation values of interest. 
We observe that under the reflection transformation $\mu \to -\mu$ and $\theta \to -\theta$, there is a quasi-symmetry exhibited by these observables: the expected value of the $U(1)$ charge and the electric field change sign $\langle Q\rangle\to -\langle Q\rangle$, $\mathcal{E}\to -\mathcal{E}$ while the chiral condensate remains invariant $\chi\to\chi$. However, we emphasize that this symmetry is not exact. 
Moreover, a striking feature we uncover is a clear hierarchical structure, where the eigenvalues of $Q$ differ by $\pm 1$ between adjacent phases. 

Subsequently, we delve into the behavior of the three aforementioned physical quantities—the U(1) charge, chiral condensate, and electric field—under variations of mass and the \(\theta\) angle, as illustrated in Fig.~\ref{fig:phase}~(d), (e), and (f). Here, we fix the chemical potential to zero, maintain a system size of \(N = 10\), and set \(a = 1/g\).
Our results reveal that the phase diagram boundaries exhibit central symmetry around the point \((\theta/(2\pi), m/g) = (-0.2, 0)\). More precisely, under the transformation \(\theta \rightarrow -\theta - 0.8\pi\) and \(m \rightarrow -m\), we find that $\langle Q\rangle\to -\langle Q\rangle$, $\chi\to -\chi$ and $\mathcal{E}\to -\mathcal{E}$.   

To deeply understand these phase diagrams, particularly the physical significance of the phase boundaries and the hierarchical structure, a thorough exploration of the phase transitions inherent in the Schwinger model is necessary.

Intuitively, when $\mu=0$ and $\theta=0$, the lowest energy levels with different $q$ present a hierarchy, since non-zero $q$ means there exists fermion or antifermion excitations and therefore have higher energy. Thus, we conjecture that
\begin{align}
    E_0^{(q)}(0,m,0)&<E_0^{(q+1)}(0,m,0), ~~& q\geq 0,\\
    E_0^{(q)}(0,m,0)&<E_0^{(q-1)}(0,m,0), ~~& q\leq 0.
\end{align}

To confirm this conjecture, we numerically computed the relationship between the lowest energy level \(E_0^{(q)}(0,m,\theta)\) and the angle \(\theta\) for various fixed charges \(q\) under zero chemical potential, as shown in Fig.~\ref{fig:energy}. Our findings confirmed that the hierarchical structure is strictly adhered to, except for two exceptional cases: when \(q < 0\) and $\theta$ close to $2\pi$, and when $q > 0$ and $\theta$ close to $-2\pi$. 

Beyond the energy level hierarchy, we observed that the curves representing the lowest energies for \(q\) and \(-q\) intersect at approximately \(\theta \approx -0.4\pi\). This observation underscores that in the absence of external fields, the equality \(E_0^{(q)}(0,m,0) = E_0^{(-q)}(0,m,0)\) does not hold, thus indicating the violation of charge conjugation symmetry. 
Indeed, a closer look at the Ref.~\cite{kokail2019self}, reveals that combined CP symmetry is preserved only in the $q=0$ state. 

The introduction of a non-zero chemical potential $\mu$ will destroy this hierarchical structure. 
Non-zero chemical potential $\mu$ induces a reconfiguration of the energy landscape, affecting the Hamiltonian's eigenvalue problem:
\begin{equation}
    H(\mu, m, \theta) \ket{q, n} = E_n^{(q)}(\mu, m,\theta) \ket{q, n},
\end{equation}
where $E_n^{(q)}(\mu, m,\theta) = E_n^{(q)}(0, m,\theta) - \mu q$ captures the energy shift introduced by the chemical potential's influence on the system's $U(1)$ charge. This shift, as elucidated in Ref.~\cite{ikeda2023detecting}, is critical for determining the $U(1)$ charge of the ground state for $H(\mu,m,\theta)$. If $E_0^{(q)}(\mu,m,\theta)$ is the ground energy, then we have 
$E_0^{(q)}(\mu, m,\theta)<E_0^{(q+1)}(\mu, m,\theta)$ and $E_0^{(q)}(\mu, m,\theta)<E_0^{(q+1)}(\mu, m,\theta)$. This implies that
\begin{equation}
    E_n^{(q)}- E_n^{(q-1)}<\mu <E_n^{(q+1)}- E_n^{(q)}. 
\end{equation}
This inequality provides us with a means to ascertain the phase boundaries.

Consider
\begin{equation}\label{eq:phase_equation}
    f^{(q)}(\mu,m,\theta)=\mu-(E_0^{(q+1)}-E_0^{(q)}),
\end{equation}
the phase boundaries between $Q=q$ and $Q=q+1$ should be the roots of the equation $f^{(q)}(\mu,m,\theta)=0$.
We can solve the equation by bisection for a given $q$, where the involved energy values are obtained through exact diagonalization. The solutions of the equations for different values of 
$q$ are in excellent agreement with the phase boundaries depicted in the phase diagrams, as shown in Fig.~\ref{fig:phase}. This result not only validates the critical role of the chemical potential in phase transitions but also attests to the accuracy of our rVQITE algorithm. 

\section{conclusion}\label{sec:conclusion}
In conclusion, this work has employed the VQITE to investigate the properties of the lattice Schwinger model. We have addressed the numerical stability issues that VQITE encounters, refining it to ensure enhanced robustness while preserving its inherent advantages.
For validation, we applied our regularized VQITE (rVQITE) to the Schwinger model at zero chemical potential and in the absence of an external field, using the Hamiltonian variational ansatz (HVA) as a variational quantum circuit. The enhanced stability and remained fast convergence of our rVQITE was demonstrated through superior performance in this benchmarking. By properly setting the initial states of the HVA, we extended our study to scenarios with finite chemical potential and external fields, presenting phase diagrams for observables of interest.

To further elucidate the phase diagram, particularly the mechanisms of phase transitions and the hierarchical structure reflected in the $U(1)$ charge, we delve into the influence of the chemical potential on the ground state $U(1)$ charge. We present the equations that delineate the phase boundaries, which we numerically solve using our rVQITE method. Our findings reveal an excellent agreement between the solutions of these equations and the phase boundaries observed in the phase diagram.

However, the quasi-symmetry in the phase diagram, which is not perfectly satisfied, lacks a comprehensive theoretical explanation, necessitating further investigation. Moreover, incorporating additional ancillary qubits or probabilistic models can extend our methodology to scenarios at finite temperatures. Additionally, exploring the extension of this work to non-Abelian gauge fields and higher spatial dimensions is a promising direction for future research endeavors.

\section*{acknowledgement}
This work was supported by the National Natural Science Foundation of China (11875160 and U1801661).

\onecolumngrid

\appendix
\section{Rrepresentation of Schwinger model}
\label{appendix:a}
Our formula framework is based on Ref.~\cite{chakraborty2022classically, nagano2023quench}.
Starting from the continuous Lagrangian density of the Schwinger model, which is given by
\begin{equation}
\mathcal{L}=-\frac{1}{4} F_{\mu \nu} F^{\mu \nu}+i \bar{\psi}\left[\gamma^\mu\left(\partial_\mu+i g A_\mu\right)-m\right] \psi+\frac{g \theta}{4 \pi} \epsilon_{\mu \nu} F^{\mu \nu},
\end{equation}
where $\psi$ represents the fermionic field, $g$ is the coupling constant, $A$ denotes the gauge field and $m$ is the fermion mass. The last term in the above expression is a topological term, which does not affect the classical equations of motion of the model but influences its quantum spectrum.
Choosing the gauge condition $A_0=0$ and introducing the canonical momentum $\Pi:=\partial \mathcal{L} / \partial\left(\partial_0 A^1\right)$, the continuous form of the Schwinger Hamiltonian can be expressed as
\begin{equation}
H=\int d x\left[\frac{1}{2}\left(\Pi-\frac{g \theta}{2 \pi}\right)^2-\mathrm{i} \bar{\psi} \gamma^1\left(\partial_1+\mathrm{i} g A_1-m\right) \psi\right],
\end{equation}
with $\Pi=\partial_0 A^1+g \theta / 2 \pi$. Due to the chosen gauge condition, Gauss's law leads to an additional constraint, requiring that any physical state $|\mathrm{phys}\rangle$ satisfies $\left(\partial_1 \Pi+g \psi^{\dagger} \psi\right)|\mathrm{phys}\rangle=0$.

In lattice gauge field theory, fermions are alternately placed on discrete lattice sites, with position $x$ discretized as $x_n, n=0,1, \cdots, N-1$. Assuming a lattice spacing $a$, we have $x_n=n a$. The fermionic field is a spinor field defined over spacetime, $\psi=\left(\psi_u(x), \psi_d(x)\right)^T$. Denoting the fermion field on the lattice site as $\chi_n$, when $n$ is odd, $\chi_n / \sqrt{a}=\psi_u\left(x_n\right)$; when $n$ is even, $\chi_n / \sqrt{a}=\psi_d(x_n)$.
These fermion field variables obey the following anticommutation relations:
\begin{align}
& \left\{\chi_n^{\dagger}, \chi_m\right\}=\delta_{m n}, \\
& \left\{\chi_n, \chi_m\right\}=0.
\end{align}
Gauge fields mediate interactions between adjacent lattice sites. Define
\begin{align}
& U_n:=e^{-i a g A^1\left(x_n\right)}, \\
& L_n:=-\Pi\left(x_n\right) / g,
\end{align}
which satisfy the relation
\begin{equation}
\begin{array}{cl}
\left[U_n, L_m\right] &= \delta_{m n} U_n \\
U_n^{\dagger} &= U_n^{-1} \\
L_n^{\dagger} &= L_n.
\end{array}
\end{equation}
Using these, we obtain the discrete Hamiltonian for the Schwinger model:
\begin{equation}
 \begin{aligned}
H &= J \sum_{n=0}^{N-2}\left(L_n+\frac{\theta}{2\pi}\right)^2\\
&- i w \sum_{n=0}^{N-2}\left(\chi_n^{\dagger} U_n \chi_{n+1} - \chi_{n+1}^{\dagger} U_n^{\dagger} \chi_n\right)\\
&+ m \sum_{n=0}^{N-1}(-1)^n \chi_n^{\dagger} \chi_n,
 \end{aligned}
\end{equation}
where $w = 1/(2a)$, $J = g^2a/2$.

The additional constraint arising from Gauss's law relates neighboring gauge field variables through
\begin{equation}
L_n - L_{n-1} = Q_n = \chi_n^{\dagger} \chi_n - \frac{\left[1 - (-1)^n\right]}{2}.
\end{equation}
By choosing appropriate boundary conditions, such as $L_{-1} = 0$, and fixing $U_n = 1$, the gauge field can be eliminated from the Hamiltonian.

Employing the Jordan-Wigner transformation,

\begin{equation}
\chi_n = \frac{\sigma_n^x - i \sigma_n^y}{2} \prod_{i=0}^{n-1}\left(-i \sigma_i^z\right),
\end{equation}

we arrive at the spin representation of the Schwinger model (up to an irrelevant constant):
\begin{equation}
 \begin{aligned}
H =& J \sum_{n=0}^{N-2}\left[\sum_{i=0}^n \frac{\sigma_i^z + (-1)^i}{2} + q\right]^2 \\
&+ \frac{w}{2} \sum_{n=0}^{N-2}\left[\sigma_n^x \sigma_{n+1}^x + \sigma_n^y \sigma_{n+1}^y\right] \\
&+ \frac{m}{2} \sum_{n=0}^{N-1}(-1)^n \sigma_n^z.
 \end{aligned}
\end{equation}
By the way, we can also derive the spin representation of $U(1)$ charge in $i$-th site
\begin{equation}
    Q_i=\frac{\sigma_i^z+(-1)^i}{2}.
\end{equation}

\section{Parameter Shift Rule for the Parameters Update}
In this appendix, we focus on the parameter shift formula for the realization of the matrix $A^R$ and the vector $C^R$ in Eq.~\eqref{eq:evolution}.

Firstly, we propose how to obtain $A^R$ and $C^R$. Consider the parameterized state $\ket{\psi(\bm{\theta})}$ has the form
\begin{equation}\label{eq:append1}  \ket{\psi(\bm{\theta})}=U_m(\theta_m)\cdots U_2(\theta_2)U_1(\theta_1)\ket{\psi_{\text{in}}},
\end{equation}
where $U_i(\theta_i)=e^{-i\theta_i g_i}$ and $g_i$ is a Pauli string.
Denote 
\begin{equation}
    C_i:=\left(\frac{\partial}{\partial \theta_i}\bra{  \psi(\bm{\theta})}\right)H\ket{\psi(\bm{\theta})},
\end{equation}
it can be expressed by 
\begin{equation}
    C_i=\bra{\psi_{\text{in}}} U_{1:i}^\dagger (ig_i)U_{i+1:m}^\dagger HU_{m:1}\ket{\psi_{in}}=:\bra{\psi_{\text{in}}}\hat{C}_i\ket{\psi_{in}},
\end{equation}
where $U_{i:j}=U_iU_{i+1}\cdots U_j$.
The Hermitian conjugation of $\hat{C}_i$ correlates the complex conjugation of $C_i$, i.e., $C_i^* = \bra{\psi_{\text{in}}}\hat{C}_i^\dagger\ket{\psi_{in}}$.
On the other hand, note that $\bra{\psi(\bm{\theta})}H\ket{ \frac{\partial}{\partial \theta_i}\psi(\bm{\theta})}=\bra{\psi_{\text{in}}}\hat{C}_i^\dagger\ket{\psi_{in}}$,
and the gradient of the energy expectation value is given by
\begin{equation}
\frac{\partial}{\partial \theta_i}\bra{\psi(\bm{\theta})}H\ket{\psi(\bm{\theta})}=\left(\frac{\partial}{\partial \theta_i}\bra{  \psi(\bm{\theta})}\right)H\ket{\psi(\bm{\theta})} +\bra{\psi(\bm{\theta})}H
\left(\frac{\partial}{\partial \theta_i}\ket{  \psi(\bm{\theta})}\right)=C_i+C_i^*,
\end{equation}
we have $C^R = \frac{1}{2}\nabla_{\bm{\theta}}\bra{\psi(\bm{\theta})}H\ket{\psi(\bm{\theta})}$. Set $\mathcal{C}(\bm{\theta}):= \bra{\psi(\bm{\theta})}H\ket{\psi(\bm{\theta})}$, following the parameter shift rule, we obtain
\begin{equation}
(C^R)_i=\left[\mathcal{C}(\bm{\theta}+\frac{\pi}{4}\bm{e}_i)-\mathcal{C}(\bm{\theta}-\frac{\pi}{4}\bm{e}_i)\right],
\end{equation}
where $\bm{e}_i = (0,\cdots,\underset{i}{1},\cdots,0)^{\text{T}}$
is the $i$-th basis in parameters space.

Now, we focus on the expression of $A^R$. Following the form of ansatz in Eq.~\eqref{eq:append1}, the matrix element of $A^R$ is given by  
\begin{equation}\label{eq:append2}
    (A^R)_{ij}=\frac{1}{2}\left[\bra{\psi(\bm{\theta}+\frac{\pi}{2}\bm{e}_i)}\ket{\psi(\bm{\theta}+\frac{\pi}{2}\bm{e}_j)}+\bra{\psi(\bm{\theta}+\frac{\pi}{2}\bm{e}_j)}\ket{\psi(\bm{\theta}+\frac{\pi}{2}\bm{e}_i)}\right].
\end{equation}
 Nonetheless, the direct measurement of the inner product implicated in the right-hand side of Eq.~\eqref{eq:append2} poses a considerable challenge. This obstacle can be circumvented by incorporating an ancillary qubit, thereby facilitating the conversion of the inner product calculation into a measurable outcome on said ancillary qubit.
Let us prepare a state with an ancilla
\begin{equation}
    \ket{\tilde{\psi}}:=\frac{1}{\sqrt{2}}\left(\ket{+}_{\text{ancilla}}\otimes \ket{\psi(\bm{\theta}+\frac{\pi}{2}\bm{e}_i)}+\ket{-}_{\text{ancilla}}\otimes\ket{\psi(\bm{\theta}+\frac{\pi}{2}\bm{e}_j)}\right), 
\end{equation}
then we have
\begin{equation}
\bra{\tilde{\psi}}\left[(\ket{0}\bra{0})_{\text{ancilla}}\otimes I\right]\ket{\tilde{\psi}}=\frac{1}{2}\left\{1+\text{Re}\left[\bra{\psi(\bm{\theta}+\frac{\pi}{2}\bm{e}_i)}\psi(\bm{\theta}+\frac{\pi}{2}\bm{e}_j)\Big\rangle\right]\right\}.
\end{equation}
Therefore, the matrix element  
\begin{equation}
    (A^R)_{ij}=2 \bra{\tilde{\psi}}[(\ket{0}\bra{0})_{\text{ancilla}}\otimes I]\ket{\tilde{\psi}}-1=\bra{\tilde{\psi}}(\sigma^z_{\text{ancilla}}\otimes I)\ket{\tilde{\psi}}.
\end{equation}


\begin{thebibliography}{32}%
\makeatletter
\providecommand \@ifxundefined [1]{%
 \@ifx{#1\undefined}
}%
\providecommand \@ifnum [1]{%
 \ifnum #1\expandafter \@firstoftwo
 \else \expandafter \@secondoftwo
 \fi
}%
\providecommand \@ifx [1]{%
 \ifx #1\expandafter \@firstoftwo
 \else \expandafter \@secondoftwo
 \fi
}%
\providecommand \natexlab [1]{#1}%
\providecommand \enquote  [1]{``#1''}%
\providecommand \bibnamefont  [1]{#1}%
\providecommand \bibfnamefont [1]{#1}%
\providecommand \citenamefont [1]{#1}%
\providecommand \href@noop [0]{\@secondoftwo}%
\providecommand \href [0]{\begingroup \@sanitize@url \@href}%
\providecommand \@href[1]{\@@startlink{#1}\@@href}%
\providecommand \@@href[1]{\endgroup#1\@@endlink}%
\providecommand \@sanitize@url [0]{\catcode `\\12\catcode `\$12\catcode `\&12\catcode `\#12\catcode `\^12\catcode `\_12\catcode `\%12\relax}%
\providecommand \@@startlink[1]{}%
\providecommand \@@endlink[0]{}%
\providecommand \url  [0]{\begingroup\@sanitize@url \@url }%
\providecommand \@url [1]{\endgroup\@href {#1}{\urlprefix }}%
\providecommand \urlprefix  [0]{URL }%
\providecommand \Eprint [0]{\href }%
\providecommand \doibase [0]{https://doi.org/}%
\providecommand \selectlanguage [0]{\@gobble}%
\providecommand \bibinfo  [0]{\@secondoftwo}%
\providecommand \bibfield  [0]{\@secondoftwo}%
\providecommand \translation [1]{[#1]}%
\providecommand \BibitemOpen [0]{}%
\providecommand \bibitemStop [0]{}%
\providecommand \bibitemNoStop [0]{.\EOS\space}%
\providecommand \EOS [0]{\spacefactor3000\relax}%
\providecommand \BibitemShut  [1]{\csname bibitem#1\endcsname}%
\let\auto@bib@innerbib\@empty
\bibitem [{\citenamefont {Martinez}\ \emph {et~al.}(2016)\citenamefont {Martinez}, \citenamefont {Muschik}, \citenamefont {Schindler}, \citenamefont {Nigg}, \citenamefont {Erhard}, \citenamefont {Heyl}, \citenamefont {Hauke}, \citenamefont {Dalmonte}, \citenamefont {Monz}, \citenamefont {Zoller} \emph {et~al.}}]{martinez2016real}%
  \BibitemOpen
  \bibfield  {author} {\bibinfo {author} {\bibfnamefont {E.~A.}\ \bibnamefont {Martinez}}, \bibinfo {author} {\bibfnamefont {C.~A.}\ \bibnamefont {Muschik}}, \bibinfo {author} {\bibfnamefont {P.}~\bibnamefont {Schindler}}, \bibinfo {author} {\bibfnamefont {D.}~\bibnamefont {Nigg}}, \bibinfo {author} {\bibfnamefont {A.}~\bibnamefont {Erhard}}, \bibinfo {author} {\bibfnamefont {M.}~\bibnamefont {Heyl}}, \bibinfo {author} {\bibfnamefont {P.}~\bibnamefont {Hauke}}, \bibinfo {author} {\bibfnamefont {M.}~\bibnamefont {Dalmonte}}, \bibinfo {author} {\bibfnamefont {T.}~\bibnamefont {Monz}}, \bibinfo {author} {\bibfnamefont {P.}~\bibnamefont {Zoller}}, \emph {et~al.},\ }\bibfield  {title} {\bibinfo {title} {Real-time dynamics of lattice gauge theories with a few-qubit quantum computer},\ }\href {https://doi.org/https://doi.org/10.1038/nature18318} {\bibfield  {journal} {\bibinfo  {journal} {Nature}\ }\textbf {\bibinfo {volume} {534}},\ \bibinfo {pages} {516} (\bibinfo {year} {2016})}\BibitemShut {NoStop}%
\bibitem [{\citenamefont {Kokail}\ \emph {et~al.}(2019)\citenamefont {Kokail}, \citenamefont {Maier}, \citenamefont {van Bijnen}, \citenamefont {Brydges}, \citenamefont {Joshi}, \citenamefont {Jurcevic}, \citenamefont {Muschik}, \citenamefont {Silvi}, \citenamefont {Blatt}, \citenamefont {Roos} \emph {et~al.}}]{kokail2019self}%
  \BibitemOpen
  \bibfield  {author} {\bibinfo {author} {\bibfnamefont {C.}~\bibnamefont {Kokail}}, \bibinfo {author} {\bibfnamefont {C.}~\bibnamefont {Maier}}, \bibinfo {author} {\bibfnamefont {R.}~\bibnamefont {van Bijnen}}, \bibinfo {author} {\bibfnamefont {T.}~\bibnamefont {Brydges}}, \bibinfo {author} {\bibfnamefont {M.~K.}\ \bibnamefont {Joshi}}, \bibinfo {author} {\bibfnamefont {P.}~\bibnamefont {Jurcevic}}, \bibinfo {author} {\bibfnamefont {C.~A.}\ \bibnamefont {Muschik}}, \bibinfo {author} {\bibfnamefont {P.}~\bibnamefont {Silvi}}, \bibinfo {author} {\bibfnamefont {R.}~\bibnamefont {Blatt}}, \bibinfo {author} {\bibfnamefont {C.~F.}\ \bibnamefont {Roos}}, \emph {et~al.},\ }\bibfield  {title} {\bibinfo {title} {Self-verifying variational quantum simulation of lattice models},\ }\href {https://doi.org/https://doi.org/10.1038/s41586-019-1177-4} {\bibfield  {journal} {\bibinfo  {journal} {Nature}\ }\textbf {\bibinfo {volume} {569}},\ \bibinfo {pages} {355} (\bibinfo {year} {2019})}\BibitemShut {NoStop}%
\bibitem [{\citenamefont {Yang}\ \emph {et~al.}(2020{\natexlab{a}})\citenamefont {Yang}, \citenamefont {Liu}, \citenamefont {Zhu}, \citenamefont {Luo},\ and\ \citenamefont {Cai}}]{yang2020simulating}%
  \BibitemOpen
  \bibfield  {author} {\bibinfo {author} {\bibfnamefont {R.-Q.}\ \bibnamefont {Yang}}, \bibinfo {author} {\bibfnamefont {H.}~\bibnamefont {Liu}}, \bibinfo {author} {\bibfnamefont {S.}~\bibnamefont {Zhu}}, \bibinfo {author} {\bibfnamefont {L.}~\bibnamefont {Luo}},\ and\ \bibinfo {author} {\bibfnamefont {R.-G.}\ \bibnamefont {Cai}},\ }\bibfield  {title} {\bibinfo {title} {Simulating quantum field theory in curved spacetime with quantum many-body systems},\ }\href {https://doi.org/10.1103/PhysRevResearch.2.023107} {\bibfield  {journal} {\bibinfo  {journal} {Phys. Rev. Res.}\ }\textbf {\bibinfo {volume} {2}},\ \bibinfo {pages} {023107} (\bibinfo {year} {2020}{\natexlab{a}})}\BibitemShut {NoStop}%
\bibitem [{\citenamefont {Yang}\ \emph {et~al.}(2020{\natexlab{b}})\citenamefont {Yang}, \citenamefont {Sun}, \citenamefont {Ott}, \citenamefont {Wang}, \citenamefont {Zache}, \citenamefont {Halimeh}, \citenamefont {Yuan}, \citenamefont {Hauke},\ and\ \citenamefont {Pan}}]{yang2020observation}%
  \BibitemOpen
  \bibfield  {author} {\bibinfo {author} {\bibfnamefont {B.}~\bibnamefont {Yang}}, \bibinfo {author} {\bibfnamefont {H.}~\bibnamefont {Sun}}, \bibinfo {author} {\bibfnamefont {R.}~\bibnamefont {Ott}}, \bibinfo {author} {\bibfnamefont {H.-Y.}\ \bibnamefont {Wang}}, \bibinfo {author} {\bibfnamefont {T.~V.}\ \bibnamefont {Zache}}, \bibinfo {author} {\bibfnamefont {J.~C.}\ \bibnamefont {Halimeh}}, \bibinfo {author} {\bibfnamefont {Z.-S.}\ \bibnamefont {Yuan}}, \bibinfo {author} {\bibfnamefont {P.}~\bibnamefont {Hauke}},\ and\ \bibinfo {author} {\bibfnamefont {J.-W.}\ \bibnamefont {Pan}},\ }\bibfield  {title} {\bibinfo {title} {Observation of gauge invariance in a 71-site bose--hubbard quantum simulator},\ }\href {https://www.nature.com/articles/s41586-020-2910-8} {\bibfield  {journal} {\bibinfo  {journal} {Nature}\ }\textbf {\bibinfo {volume} {587}},\ \bibinfo {pages} {392} (\bibinfo {year} {2020}{\natexlab{b}})}\BibitemShut {NoStop}%
\bibitem [{\citenamefont {Atas}\ \emph {et~al.}(2021)\citenamefont {Atas}, \citenamefont {Zhang}, \citenamefont {Lewis}, \citenamefont {Jahanpour}, \citenamefont {Haase},\ and\ \citenamefont {Muschik}}]{atas20212su}%
  \BibitemOpen
  \bibfield  {author} {\bibinfo {author} {\bibfnamefont {Y.~Y.}\ \bibnamefont {Atas}}, \bibinfo {author} {\bibfnamefont {J.}~\bibnamefont {Zhang}}, \bibinfo {author} {\bibfnamefont {R.}~\bibnamefont {Lewis}}, \bibinfo {author} {\bibfnamefont {A.}~\bibnamefont {Jahanpour}}, \bibinfo {author} {\bibfnamefont {J.~F.}\ \bibnamefont {Haase}},\ and\ \bibinfo {author} {\bibfnamefont {C.~A.}\ \bibnamefont {Muschik}},\ }\bibfield  {title} {\bibinfo {title} {Su (2) hadrons on a quantum computer via a variational approach},\ }\href {https://doi.org/https://doi.org/10.1038/s41467-021-26825-4} {\bibfield  {journal} {\bibinfo  {journal} {Nature communications}\ }\textbf {\bibinfo {volume} {12}},\ \bibinfo {pages} {6499} (\bibinfo {year} {2021})}\BibitemShut {NoStop}%
\bibitem [{\citenamefont {Chakraborty}\ \emph {et~al.}(2022)\citenamefont {Chakraborty}, \citenamefont {Honda}, \citenamefont {Izubuchi}, \citenamefont {Kikuchi},\ and\ \citenamefont {Tomiya}}]{chakraborty2022classically}%
  \BibitemOpen
  \bibfield  {author} {\bibinfo {author} {\bibfnamefont {B.}~\bibnamefont {Chakraborty}}, \bibinfo {author} {\bibfnamefont {M.}~\bibnamefont {Honda}}, \bibinfo {author} {\bibfnamefont {T.}~\bibnamefont {Izubuchi}}, \bibinfo {author} {\bibfnamefont {Y.}~\bibnamefont {Kikuchi}},\ and\ \bibinfo {author} {\bibfnamefont {A.}~\bibnamefont {Tomiya}},\ }\bibfield  {title} {\bibinfo {title} {Classically emulated digital quantum simulation of the schwinger model with a topological term via adiabatic state preparation},\ }\href {https://doi.org/10.1103/PhysRevD.105.094503} {\bibfield  {journal} {\bibinfo  {journal} {Phys. Rev. D}\ }\textbf {\bibinfo {volume} {105}},\ \bibinfo {pages} {094503} (\bibinfo {year} {2022})}\BibitemShut {NoStop}%
\bibitem [{\citenamefont {Nagano}\ \emph {et~al.}(2023)\citenamefont {Nagano}, \citenamefont {Bapat},\ and\ \citenamefont {Bauer}}]{nagano2023quench}%
  \BibitemOpen
  \bibfield  {author} {\bibinfo {author} {\bibfnamefont {L.}~\bibnamefont {Nagano}}, \bibinfo {author} {\bibfnamefont {A.}~\bibnamefont {Bapat}},\ and\ \bibinfo {author} {\bibfnamefont {C.~W.}\ \bibnamefont {Bauer}},\ }\bibfield  {title} {\bibinfo {title} {Quench dynamics of the schwinger model via variational quantum algorithms},\ }\href {https://doi.org/10.1103/PhysRevD.108.034501} {\bibfield  {journal} {\bibinfo  {journal} {Phys. Rev. D}\ }\textbf {\bibinfo {volume} {108}},\ \bibinfo {pages} {034501} (\bibinfo {year} {2023})}\BibitemShut {NoStop}%
\bibitem [{\citenamefont {Ikeda}\ \emph {et~al.}(2023)\citenamefont {Ikeda}, \citenamefont {Kharzeev}, \citenamefont {Meyer},\ and\ \citenamefont {Shi}}]{ikeda2023detecting}%
  \BibitemOpen
  \bibfield  {author} {\bibinfo {author} {\bibfnamefont {K.}~\bibnamefont {Ikeda}}, \bibinfo {author} {\bibfnamefont {D.~E.}\ \bibnamefont {Kharzeev}}, \bibinfo {author} {\bibfnamefont {R.}~\bibnamefont {Meyer}},\ and\ \bibinfo {author} {\bibfnamefont {S.}~\bibnamefont {Shi}},\ }\bibfield  {title} {\bibinfo {title} {Detecting the critical point through entanglement in the schwinger model},\ }\href {https://doi.org/10.1103/PhysRevD.108.L091501} {\bibfield  {journal} {\bibinfo  {journal} {Phys. Rev. D}\ }\textbf {\bibinfo {volume} {108}},\ \bibinfo {pages} {L091501} (\bibinfo {year} {2023})}\BibitemShut {NoStop}%
\bibitem [{\citenamefont {Schwinger}(1962)}]{schwinger1962gauge}%
  \BibitemOpen
  \bibfield  {author} {\bibinfo {author} {\bibfnamefont {J.}~\bibnamefont {Schwinger}},\ }\bibfield  {title} {\bibinfo {title} {Gauge invariance and mass. {II}},\ }\href {https://doi.org/10.1103/PhysRev.128.2425} {\bibfield  {journal} {\bibinfo  {journal} {Phys. Rev.}\ }\textbf {\bibinfo {volume} {128}},\ \bibinfo {pages} {2425} (\bibinfo {year} {1962})}\BibitemShut {NoStop}%
\bibitem [{\citenamefont {Lowenstein}\ and\ \citenamefont {Swieca}(1971)}]{lowenstein1971quantum}%
  \BibitemOpen
  \bibfield  {author} {\bibinfo {author} {\bibfnamefont {J.}~\bibnamefont {Lowenstein}}\ and\ \bibinfo {author} {\bibfnamefont {J.}~\bibnamefont {Swieca}},\ }\bibfield  {title} {\bibinfo {title} {Quantum electrodynamics in two dimensions},\ }\href {https://doi.org/https://doi.org/10.1016/0003-4916(71)90246-6} {\bibfield  {journal} {\bibinfo  {journal} {Annals of Physics}\ }\textbf {\bibinfo {volume} {68}},\ \bibinfo {pages} {172} (\bibinfo {year} {1971})}\BibitemShut {NoStop}%
\bibitem [{\citenamefont {Coleman}\ \emph {et~al.}(1975)\citenamefont {Coleman}, \citenamefont {Jackiw},\ and\ \citenamefont {Susskind}}]{coleman1975charge}%
  \BibitemOpen
  \bibfield  {author} {\bibinfo {author} {\bibfnamefont {S.}~\bibnamefont {Coleman}}, \bibinfo {author} {\bibfnamefont {R.}~\bibnamefont {Jackiw}},\ and\ \bibinfo {author} {\bibfnamefont {L.}~\bibnamefont {Susskind}},\ }\bibfield  {title} {\bibinfo {title} {Charge shielding and quark confinement in the massive schwinger model},\ }\href {https://doi.org/https://doi.org/10.1016/0003-4916(75)90212-2} {\bibfield  {journal} {\bibinfo  {journal} {Annals of Physics}\ }\textbf {\bibinfo {volume} {93}},\ \bibinfo {pages} {267} (\bibinfo {year} {1975})}\BibitemShut {NoStop}%
\bibitem [{\citenamefont {Gross}\ \emph {et~al.}(1996)\citenamefont {Gross}, \citenamefont {Klebanov}, \citenamefont {Matytsin},\ and\ \citenamefont {Smilga}}]{gross1996screening}%
  \BibitemOpen
  \bibfield  {author} {\bibinfo {author} {\bibfnamefont {D.~J.}\ \bibnamefont {Gross}}, \bibinfo {author} {\bibfnamefont {I.~R.}\ \bibnamefont {Klebanov}}, \bibinfo {author} {\bibfnamefont {A.~V.}\ \bibnamefont {Matytsin}},\ and\ \bibinfo {author} {\bibfnamefont {A.~V.}\ \bibnamefont {Smilga}},\ }\bibfield  {title} {\bibinfo {title} {Screening versus confinement in 1 + 1 dimensions},\ }\href {https://doi.org/https://doi.org/10.1016/0550-3213(95)00655-9} {\bibfield  {journal} {\bibinfo  {journal} {Nuclear Physics B}\ }\textbf {\bibinfo {volume} {461}},\ \bibinfo {pages} {109} (\bibinfo {year} {1996})}\BibitemShut {NoStop}%
\bibitem [{\citenamefont {Troyer}\ and\ \citenamefont {Wiese}(2005)}]{troyer2005computational}%
  \BibitemOpen
  \bibfield  {author} {\bibinfo {author} {\bibfnamefont {M.}~\bibnamefont {Troyer}}\ and\ \bibinfo {author} {\bibfnamefont {U.-J.}\ \bibnamefont {Wiese}},\ }\bibfield  {title} {\bibinfo {title} {Computational complexity and fundamental limitations to fermionic quantum monte carlo simulations},\ }\href {https://doi.org/10.1103/PhysRevLett.94.170201} {\bibfield  {journal} {\bibinfo  {journal} {Phys. Rev. Lett.}\ }\textbf {\bibinfo {volume} {94}},\ \bibinfo {pages} {170201} (\bibinfo {year} {2005})}\BibitemShut {NoStop}%
\bibitem [{\citenamefont {Fukushima}\ and\ \citenamefont {Hatsuda}(2010)}]{fukushima2010phase}%
  \BibitemOpen
  \bibfield  {author} {\bibinfo {author} {\bibfnamefont {K.}~\bibnamefont {Fukushima}}\ and\ \bibinfo {author} {\bibfnamefont {T.}~\bibnamefont {Hatsuda}},\ }\bibfield  {title} {\bibinfo {title} {The phase diagram of dense {QCD}},\ }\href {https://doi.org/10.1088/0034-4885/74/1/014001} {\bibfield  {journal} {\bibinfo  {journal} {Reports on Progress in Physics}\ }\textbf {\bibinfo {volume} {74}},\ \bibinfo {pages} {014001} (\bibinfo {year} {2010})}\BibitemShut {NoStop}%
\bibitem [{\citenamefont {Nagata}(2022)}]{nagata2022finite}%
  \BibitemOpen
  \bibfield  {author} {\bibinfo {author} {\bibfnamefont {K.}~\bibnamefont {Nagata}},\ }\bibfield  {title} {\bibinfo {title} {Finite-density lattice {QCD} and sign problem: Current status and open problems},\ }\href {https://doi.org/https://doi.org/10.1016/j.ppnp.2022.103991} {\bibfield  {journal} {\bibinfo  {journal} {Progress in Particle and Nuclear Physics}\ }\textbf {\bibinfo {volume} {127}},\ \bibinfo {pages} {103991} (\bibinfo {year} {2022})}\BibitemShut {NoStop}%
\bibitem [{\citenamefont {Kogut}\ and\ \citenamefont {Susskind}(1975)}]{kogut1975hamiltonian}%
  \BibitemOpen
  \bibfield  {author} {\bibinfo {author} {\bibfnamefont {J.}~\bibnamefont {Kogut}}\ and\ \bibinfo {author} {\bibfnamefont {L.}~\bibnamefont {Susskind}},\ }\bibfield  {title} {\bibinfo {title} {Hamiltonian formulation of wilson's lattice gauge theories},\ }\href {https://doi.org/10.1103/PhysRevD.11.395} {\bibfield  {journal} {\bibinfo  {journal} {Phys. Rev. D}\ }\textbf {\bibinfo {volume} {11}},\ \bibinfo {pages} {395} (\bibinfo {year} {1975})}\BibitemShut {NoStop}%
\bibitem [{\citenamefont {Tomiya}(2022)}]{tomiya2022schwinger}%
  \BibitemOpen
  \bibfield  {author} {\bibinfo {author} {\bibfnamefont {A.}~\bibnamefont {Tomiya}},\ }\bibfield  {title} {\bibinfo {title} {Schwinger model at finite temperature and density with beta vqe},\ }\href {https://doi.org/10.48550/arXiv.2205.08860} {\bibfield  {journal} {\bibinfo  {journal} {arXiv:2205.08860}\ } (\bibinfo {year} {2022})}\BibitemShut {NoStop}%
\bibitem [{\citenamefont {McArdle}\ \emph {et~al.}(2019)\citenamefont {McArdle}, \citenamefont {Jones}, \citenamefont {Endo}, \citenamefont {Li}, \citenamefont {Benjamin},\ and\ \citenamefont {Yuan}}]{mcardle2019variational}%
  \BibitemOpen
  \bibfield  {author} {\bibinfo {author} {\bibfnamefont {S.}~\bibnamefont {McArdle}}, \bibinfo {author} {\bibfnamefont {T.}~\bibnamefont {Jones}}, \bibinfo {author} {\bibfnamefont {S.}~\bibnamefont {Endo}}, \bibinfo {author} {\bibfnamefont {Y.}~\bibnamefont {Li}}, \bibinfo {author} {\bibfnamefont {S.~C.}\ \bibnamefont {Benjamin}},\ and\ \bibinfo {author} {\bibfnamefont {X.}~\bibnamefont {Yuan}},\ }\bibfield  {title} {\bibinfo {title} {Variational ansatz-based quantum simulation of imaginary time evolution},\ }\href {https://www.nature.com/articles/s41534-019-0187-2} {\bibfield  {journal} {\bibinfo  {journal} {npj Quantum Information}\ }\textbf {\bibinfo {volume} {5}},\ \bibinfo {pages} {75} (\bibinfo {year} {2019})}\BibitemShut {NoStop}%
\bibitem [{\citenamefont {Motta}\ \emph {et~al.}(2020)\citenamefont {Motta}, \citenamefont {Sun}, \citenamefont {Tan}, \citenamefont {O’Rourke}, \citenamefont {Ye}, \citenamefont {Minnich}, \citenamefont {Brandao},\ and\ \citenamefont {Chan}}]{motta2020determining}%
  \BibitemOpen
  \bibfield  {author} {\bibinfo {author} {\bibfnamefont {M.}~\bibnamefont {Motta}}, \bibinfo {author} {\bibfnamefont {C.}~\bibnamefont {Sun}}, \bibinfo {author} {\bibfnamefont {A.~T.}\ \bibnamefont {Tan}}, \bibinfo {author} {\bibfnamefont {M.~J.}\ \bibnamefont {O’Rourke}}, \bibinfo {author} {\bibfnamefont {E.}~\bibnamefont {Ye}}, \bibinfo {author} {\bibfnamefont {A.~J.}\ \bibnamefont {Minnich}}, \bibinfo {author} {\bibfnamefont {F.~G.}\ \bibnamefont {Brandao}},\ and\ \bibinfo {author} {\bibfnamefont {G.~K.-L.}\ \bibnamefont {Chan}},\ }\bibfield  {title} {\bibinfo {title} {Determining eigenstates and thermal states on a quantum computer using quantum imaginary time evolution},\ }\href {https://www.nature.com/articles/s41567-019-0704-4} {\bibfield  {journal} {\bibinfo  {journal} {Nature Physics}\ }\textbf {\bibinfo {volume} {16}},\ \bibinfo {pages} {205} (\bibinfo {year} {2020})}\BibitemShut {NoStop}%
\bibitem [{\citenamefont {Yeter-Aydeniz}\ \emph {et~al.}(2022)\citenamefont {Yeter-Aydeniz}, \citenamefont {Moschandreou},\ and\ \citenamefont {Siopsis}}]{yeter2022quantum}%
  \BibitemOpen
  \bibfield  {author} {\bibinfo {author} {\bibfnamefont {K.}~\bibnamefont {Yeter-Aydeniz}}, \bibinfo {author} {\bibfnamefont {E.}~\bibnamefont {Moschandreou}},\ and\ \bibinfo {author} {\bibfnamefont {G.}~\bibnamefont {Siopsis}},\ }\bibfield  {title} {\bibinfo {title} {Quantum imaginary-time evolution algorithm for quantum field theories with continuous variables},\ }\href {https://doi.org/10.1103/PhysRevA.105.012412} {\bibfield  {journal} {\bibinfo  {journal} {Phys. Rev. A}\ }\textbf {\bibinfo {volume} {105}},\ \bibinfo {pages} {012412} (\bibinfo {year} {2022})}\BibitemShut {NoStop}%
\bibitem [{\citenamefont {Yuan}\ \emph {et~al.}(2019)\citenamefont {Yuan}, \citenamefont {Endo}, \citenamefont {Zhao}, \citenamefont {Li},\ and\ \citenamefont {Benjamin}}]{Yuan2019theoryofvariational}%
  \BibitemOpen
  \bibfield  {author} {\bibinfo {author} {\bibfnamefont {X.}~\bibnamefont {Yuan}}, \bibinfo {author} {\bibfnamefont {S.}~\bibnamefont {Endo}}, \bibinfo {author} {\bibfnamefont {Q.}~\bibnamefont {Zhao}}, \bibinfo {author} {\bibfnamefont {Y.}~\bibnamefont {Li}},\ and\ \bibinfo {author} {\bibfnamefont {S.~C.}\ \bibnamefont {Benjamin}},\ }\bibfield  {title} {\bibinfo {title} {Theory of variational quantum simulation},\ }\href {https://doi.org/10.22331/q-2019-10-07-191} {\bibfield  {journal} {\bibinfo  {journal} {{Quantum}}\ }\textbf {\bibinfo {volume} {3}},\ \bibinfo {pages} {191} (\bibinfo {year} {2019})}\BibitemShut {NoStop}%
\bibitem [{\citenamefont {Liu}\ \emph {et~al.}(2021)\citenamefont {Liu}, \citenamefont {Liu},\ and\ \citenamefont {Fan}}]{liu2021probabilistic}%
  \BibitemOpen
  \bibfield  {author} {\bibinfo {author} {\bibfnamefont {T.}~\bibnamefont {Liu}}, \bibinfo {author} {\bibfnamefont {J.-G.}\ \bibnamefont {Liu}},\ and\ \bibinfo {author} {\bibfnamefont {H.}~\bibnamefont {Fan}},\ }\bibfield  {title} {\bibinfo {title} {Probabilistic nonunitary gate in imaginary time evolution},\ }\href {https://doi.org/https://doi.org/10.1007/s11128-021-03145-6} {\bibfield  {journal} {\bibinfo  {journal} {Quantum Information Processing}\ }\textbf {\bibinfo {volume} {20}},\ \bibinfo {pages} {204} (\bibinfo {year} {2021})}\BibitemShut {NoStop}%
\bibitem [{\citenamefont {Kosugi}\ \emph {et~al.}(2021)\citenamefont {Kosugi}, \citenamefont {Nishiya}, \citenamefont {Nishi},\ and\ \citenamefont {Matsushita}}]{kosugi2021probabilistic}%
  \BibitemOpen
  \bibfield  {author} {\bibinfo {author} {\bibfnamefont {T.}~\bibnamefont {Kosugi}}, \bibinfo {author} {\bibfnamefont {Y.}~\bibnamefont {Nishiya}}, \bibinfo {author} {\bibfnamefont {H.}~\bibnamefont {Nishi}},\ and\ \bibinfo {author} {\bibfnamefont {Y.-i.}\ \bibnamefont {Matsushita}},\ }\bibfield  {title} {\bibinfo {title} {Probabilistic imaginary-time evolution by using forward and backward real-time evolution with a single ancilla: first-quantized eigensolver of quantum chemistry for ground states},\ }\href {https://arxiv.org/abs/2111.12471} {\bibfield  {journal} {\bibinfo  {journal} {arXiv:2111.12471}\ } (\bibinfo {year} {2021})}\BibitemShut {NoStop}%
\bibitem [{\citenamefont {Xie}\ \emph {et~al.}(2024)\citenamefont {Xie}, \citenamefont {Wei}, \citenamefont {Yang}, \citenamefont {Wang}, \citenamefont {Chen}, \citenamefont {Fan},\ and\ \citenamefont {Long}}]{xie2024probabilistic}%
  \BibitemOpen
  \bibfield  {author} {\bibinfo {author} {\bibfnamefont {H.-N.}\ \bibnamefont {Xie}}, \bibinfo {author} {\bibfnamefont {S.-J.}\ \bibnamefont {Wei}}, \bibinfo {author} {\bibfnamefont {F.}~\bibnamefont {Yang}}, \bibinfo {author} {\bibfnamefont {Z.-A.}\ \bibnamefont {Wang}}, \bibinfo {author} {\bibfnamefont {C.-T.}\ \bibnamefont {Chen}}, \bibinfo {author} {\bibfnamefont {H.}~\bibnamefont {Fan}},\ and\ \bibinfo {author} {\bibfnamefont {G.-L.}\ \bibnamefont {Long}},\ }\bibfield  {title} {\bibinfo {title} {Probabilistic imaginary-time evolution algorithm based on nonunitary quantum circuits},\ }\href {https://doi.org/10.1103/PhysRevA.109.052414} {\bibfield  {journal} {\bibinfo  {journal} {Phys. Rev. A}\ }\textbf {\bibinfo {volume} {109}},\ \bibinfo {pages} {052414} (\bibinfo {year} {2024})}\BibitemShut {NoStop}%
\bibitem [{\citenamefont {Wiersema}\ \emph {et~al.}(2020)\citenamefont {Wiersema}, \citenamefont {Zhou}, \citenamefont {de~Sereville}, \citenamefont {Carrasquilla}, \citenamefont {Kim},\ and\ \citenamefont {Yuen}}]{wiersema2020exploring}%
  \BibitemOpen
  \bibfield  {author} {\bibinfo {author} {\bibfnamefont {R.}~\bibnamefont {Wiersema}}, \bibinfo {author} {\bibfnamefont {C.}~\bibnamefont {Zhou}}, \bibinfo {author} {\bibfnamefont {Y.}~\bibnamefont {de~Sereville}}, \bibinfo {author} {\bibfnamefont {J.~F.}\ \bibnamefont {Carrasquilla}}, \bibinfo {author} {\bibfnamefont {Y.~B.}\ \bibnamefont {Kim}},\ and\ \bibinfo {author} {\bibfnamefont {H.}~\bibnamefont {Yuen}},\ }\bibfield  {title} {\bibinfo {title} {Exploring entanglement and optimization within the hamiltonian variational ansatz},\ }\href {https://doi.org/10.1103/PRXQuantum.1.020319} {\bibfield  {journal} {\bibinfo  {journal} {PRX Quantum}\ }\textbf {\bibinfo {volume} {1}},\ \bibinfo {pages} {020319} (\bibinfo {year} {2020})}\BibitemShut {NoStop}%
\bibitem [{\citenamefont {Beach}\ \emph {et~al.}(2019)\citenamefont {Beach}, \citenamefont {Melko}, \citenamefont {Grover},\ and\ \citenamefont {Hsieh}}]{bench2019making}%
  \BibitemOpen
  \bibfield  {author} {\bibinfo {author} {\bibfnamefont {M.~J.~S.}\ \bibnamefont {Beach}}, \bibinfo {author} {\bibfnamefont {R.~G.}\ \bibnamefont {Melko}}, \bibinfo {author} {\bibfnamefont {T.}~\bibnamefont {Grover}},\ and\ \bibinfo {author} {\bibfnamefont {T.~H.}\ \bibnamefont {Hsieh}},\ }\bibfield  {title} {\bibinfo {title} {Making trotters sprint: A variational imaginary time ansatz for quantum many-body systems},\ }\href {https://doi.org/10.1103/PhysRevB.100.094434} {\bibfield  {journal} {\bibinfo  {journal} {Phys. Rev. B}\ }\textbf {\bibinfo {volume} {100}},\ \bibinfo {pages} {094434} (\bibinfo {year} {2019})}\BibitemShut {NoStop}%
\bibitem [{\citenamefont {Gomes}\ \emph {et~al.}(2021)\citenamefont {Gomes}, \citenamefont {Mukherjee}, \citenamefont {Zhang}, \citenamefont {Iadecola}, \citenamefont {Wang}, \citenamefont {Ho}, \citenamefont {Orth},\ and\ \citenamefont {Yao}}]{gomes2021adaptive}%
  \BibitemOpen
  \bibfield  {author} {\bibinfo {author} {\bibfnamefont {N.}~\bibnamefont {Gomes}}, \bibinfo {author} {\bibfnamefont {A.}~\bibnamefont {Mukherjee}}, \bibinfo {author} {\bibfnamefont {F.}~\bibnamefont {Zhang}}, \bibinfo {author} {\bibfnamefont {T.}~\bibnamefont {Iadecola}}, \bibinfo {author} {\bibfnamefont {C.-Z.}\ \bibnamefont {Wang}}, \bibinfo {author} {\bibfnamefont {K.-M.}\ \bibnamefont {Ho}}, \bibinfo {author} {\bibfnamefont {P.~P.}\ \bibnamefont {Orth}},\ and\ \bibinfo {author} {\bibfnamefont {Y.-X.}\ \bibnamefont {Yao}},\ }\bibfield  {title} {\bibinfo {title} {Adaptive variational quantum imaginary time evolution approach for ground state preparation},\ }\href {https://doi.org/https://doi.org/10.1002/qute.202100114} {\bibfield  {journal} {\bibinfo  {journal} {Advanced Quantum Technologies}\ }\textbf {\bibinfo {volume} {4}},\ \bibinfo {pages} {2100114} (\bibinfo {year} {2021})}\BibitemShut {NoStop}%
\bibitem [{\citenamefont {Stokes}\ \emph {et~al.}(2020)\citenamefont {Stokes}, \citenamefont {Izaac}, \citenamefont {Killoran},\ and\ \citenamefont {Carleo}}]{stokes2020quantumnatural}%
  \BibitemOpen
  \bibfield  {author} {\bibinfo {author} {\bibfnamefont {J.}~\bibnamefont {Stokes}}, \bibinfo {author} {\bibfnamefont {J.}~\bibnamefont {Izaac}}, \bibinfo {author} {\bibfnamefont {N.}~\bibnamefont {Killoran}},\ and\ \bibinfo {author} {\bibfnamefont {G.}~\bibnamefont {Carleo}},\ }\bibfield  {title} {\bibinfo {title} {Quantum {N}atural {G}radient},\ }\href {https://doi.org/10.22331/q-2020-05-25-269} {\bibfield  {journal} {\bibinfo  {journal} {{Quantum}}\ }\textbf {\bibinfo {volume} {4}},\ \bibinfo {pages} {269} (\bibinfo {year} {2020})}\BibitemShut {NoStop}%
\bibitem [{\citenamefont {Anguas}\ \emph {et~al.}(2019)\citenamefont {Anguas}, \citenamefont {Bueno},\ and\ \citenamefont {Dopico}}]{anguas2019comparison}%
  \BibitemOpen
  \bibfield  {author} {\bibinfo {author} {\bibfnamefont {L.~M.}\ \bibnamefont {Anguas}}, \bibinfo {author} {\bibfnamefont {M.~I.}\ \bibnamefont {Bueno}},\ and\ \bibinfo {author} {\bibfnamefont {F.~M.}\ \bibnamefont {Dopico}},\ }\bibfield  {title} {\bibinfo {title} {A comparison of eigenvalue condition numbers for matrix polynomials},\ }\href {https://doi.org/https://doi.org/10.1016/j.laa.2018.11.031} {\bibfield  {journal} {\bibinfo  {journal} {Linear Algebra and its Applications}\ }\textbf {\bibinfo {volume} {564}},\ \bibinfo {pages} {170} (\bibinfo {year} {2019})}\BibitemShut {NoStop}%
\bibitem [{\citenamefont {{El Ghaoui}}(2002)}]{el2002inversion}%
  \BibitemOpen
  \bibfield  {author} {\bibinfo {author} {\bibfnamefont {L.}~\bibnamefont {{El Ghaoui}}},\ }\bibfield  {title} {\bibinfo {title} {Inversion error, condition number, and approximate inverses of uncertain matrices},\ }\href {https://doi.org/https://doi.org/10.1016/S0024-3795(01)00273-7} {\bibfield  {journal} {\bibinfo  {journal} {Linear Algebra and its Applications}\ }\textbf {\bibinfo {volume} {343-344}},\ \bibinfo {pages} {171} (\bibinfo {year} {2002})},\ \bibinfo {note} {special Issue on Structured and Infinite Systems of Linear equations}\BibitemShut {NoStop}%
\bibitem [{\citenamefont {Developer}(2021)}]{mq_2021}%
  \BibitemOpen
  \bibfield  {author} {\bibinfo {author} {\bibfnamefont {M.}~\bibnamefont {Developer}},\ }\href {https://gitee.com/mindspore/mindquantum} {\bibinfo {title} {Mindquantum, version 0.6.0}} (\bibinfo {year} {2021})\BibitemShut {NoStop}%
\bibitem [{\citenamefont {Pagano}\ \emph {et~al.}(2020)\citenamefont {Pagano}, \citenamefont {Bapat}, \citenamefont {Becker}, \citenamefont {Collins}, \citenamefont {De}, \citenamefont {Hess}, \citenamefont {Kaplan}, \citenamefont {Kyprianidis}, \citenamefont {Tan}, \citenamefont {Baldwin} \emph {et~al.}}]{pagano2020quantum}%
  \BibitemOpen
  \bibfield  {author} {\bibinfo {author} {\bibfnamefont {G.}~\bibnamefont {Pagano}}, \bibinfo {author} {\bibfnamefont {A.}~\bibnamefont {Bapat}}, \bibinfo {author} {\bibfnamefont {P.}~\bibnamefont {Becker}}, \bibinfo {author} {\bibfnamefont {K.~S.}\ \bibnamefont {Collins}}, \bibinfo {author} {\bibfnamefont {A.}~\bibnamefont {De}}, \bibinfo {author} {\bibfnamefont {P.~W.}\ \bibnamefont {Hess}}, \bibinfo {author} {\bibfnamefont {H.~B.}\ \bibnamefont {Kaplan}}, \bibinfo {author} {\bibfnamefont {A.}~\bibnamefont {Kyprianidis}}, \bibinfo {author} {\bibfnamefont {W.~L.}\ \bibnamefont {Tan}}, \bibinfo {author} {\bibfnamefont {C.}~\bibnamefont {Baldwin}}, \emph {et~al.},\ }\bibfield  {title} {\bibinfo {title} {Quantum approximate optimization of the long-range ising model with a trapped-ion quantum simulator},\ }\href {https://doi.org/https://doi.org/10.1073/pnas.2006373117} {\bibfield  {journal} {\bibinfo  {journal} {Proceedings of the National Academy of Sciences}\ }\textbf {\bibinfo {volume} {117}},\ \bibinfo
  {pages} {25396} (\bibinfo {year} {2020})}\BibitemShut {NoStop}%
\end{thebibliography}
\end{document}